\documentclass[aps,prb,twocolumn,showpacs,10pt,floatfix,longbibliography]{revtex4-1}

\usepackage[pdftex]{graphicx} 
\usepackage{amsmath,amssymb}
\usepackage{bm}
\usepackage{graphicx}
\usepackage{epstopdf}
\usepackage{latexsym} 
\usepackage{subfigure}
\usepackage{color}
\usepackage{dsfont} 
\usepackage{wasysym} 
\usepackage[hypertexnames=false]{hyperref}
\usepackage{verbatim}
\usepackage{tabularx}
\usepackage{amsfonts}

\newcommand{\be}{\begin{equation}}
\newcommand{\bea}{\begin{eqnarray}}
\newcommand{\ee}{\end{equation}}
\newcommand{\eea}{\end{eqnarray}}

\begin{document}
\title{Cluster Mott insulators and two Curie-Weiss regimes on an anisotropic Kagome lattice}
\author{Gang Chen$^{1,2,3,7}$}
\email{Correponding email: gchen$_$physics@fudan.edu.cn or gangchen.physics@gmail.com}
\author{Hae-Young Kee$^{4,5}$}
\author{Yong Baek Kim$^{4,5,6}$}
\affiliation{${}^1$State Key Laboratory of Surface Physics, Fudan University, Shanghai 200433, People's Republic of China}
\affiliation{${}^2$Department of Physics, Fudan University, Shanghai 200433, People's Republic of China}
\affiliation{${}^3$Center for Field Theory and Particle Physics,
Fudan University, Shanghai, 200433, People's Republic of China}
\affiliation{${}^4$Department of Physics, University of Toronto, Toronto, Ontario, M5S1A7, Canada}
\affiliation{${}^5$Canadian Institute for Advanced Research/Quantum Materials Program, 
Toronto, Ontario MSG 1Z8, Canada}
\affiliation{${}^6$School of Physics, Korea Institute for Advanced Study, Seoul 130-722, Korea}
\affiliation{${}^7$Collaborative Innovation Center of Advanced Microstructures,
Nanjing 210093, People's Republic of China}

\date{\today}

\begin{abstract}
Motivated by recent experiments on the quantum-spin-liquid candidate material 
LiZn$_2$Mo$_3$O$_8$, we study a single-band extended Hubbard model on an 
anisotropic Kagome lattice with the 1/6 electron filling. Due to the partial 
filling of the lattice, the inter-site repulsive interaction is necessary to 
generate Mott insulators, where electrons are localized in clusters, 
rather than at lattice sites. It is shown that these cluster Mott insulators 
are generally U(1) quantum spin liquids with spinon Fermi surfaces. 
The nature of charge excitations in cluster Mott insulators can be quite 
different from conventional Mott insulator and we show that there exists 
a novel cluster Mott insulator where charge fluctuations around the hexagonal 
cluster induce a plaquette charge order (PCO). The spinon excitation spectrum 
in this spin-liquid cluster Mott insulator is reconstructed due to the PCO so 
that only 1/3 of the total spinon excitations are magnetically active. Based 
on these results, we propose that the two Curie-Weiss regimes of the spin 
susceptibility in LiZn$_2$Mo$_3$O$_8$ may be explained by finite-temperature 
properties of the cluster Mott insulator with the PCO as well as fractionalized 
spinon excitations. Existing and possible future experiments on LiZn$_2$Mo$_3$O$_8$, 
and other Mo-based cluster magnets are discussed in light of these theoretical 
predictions.
\end{abstract}

\maketitle

\section{Introduction}
\label{sec1}

If there is no spontaneous symmetry breaking, the ground state of a Mott insulator with 
odd number of electrons per unit cell may be a quantum spin liquid (QSL)~\cite{Hastings04}. 
QSL is an exotic quantum phase of matter with a long-range quantum entanglement~\cite{Wenbook} 
and is characterized by fractionalized spin excitations 
and an emergent gauge structures at low energies~\cite{Lee05092008,Moessner2001,Balents10}. 
It has been suggested that some frustrated Mott insulating systems which are proximate to 
Mott transitions may provide physical realizations of QSLs~\cite{Lee05,Motrunich05,Motrunich06,Podolsky09,PhysRevB.87.165120}.
These QSLs are believed to arise 
from strong charge fluctuations in the weak Mott regime,
which can generate sizable long-range spin exchanges or spin ring exchanges 
and suppress possible magnetic orderings~\cite{Motrunich05,Motrunich06}.
Several QSL candidate materials, such as the 2D 
triangular lattice organic materials $\kappa$-(ET)$_2$Cu$_2$(CN)$_3$ 
and EtMe$_3$Sb[Pd(dmit)$_2$]$_2$,
and a 3D hyperkagome system Na$_4$Ir$_3$O$_8$~\cite{itou08,Shimizu03,Okamoto07}, 
are expected to be in this weak Mott regime.
These weak Mott-insulator U(1) QSLs are obtained as a deconfined phase of an
emergent U(1) lattice gauge theory~\cite{Florens04,Lee05}, where 
the electron is fractionalized into spin-carrying spinons and charged bosons.
The charge excitations are gapped and the low-energy physics of the QSLs is 
described by a spinon Fermi surface coupled to the emergent U(1) gauge field. 

\begin{figure}[t]
 {\includegraphics[width=8.5cm]{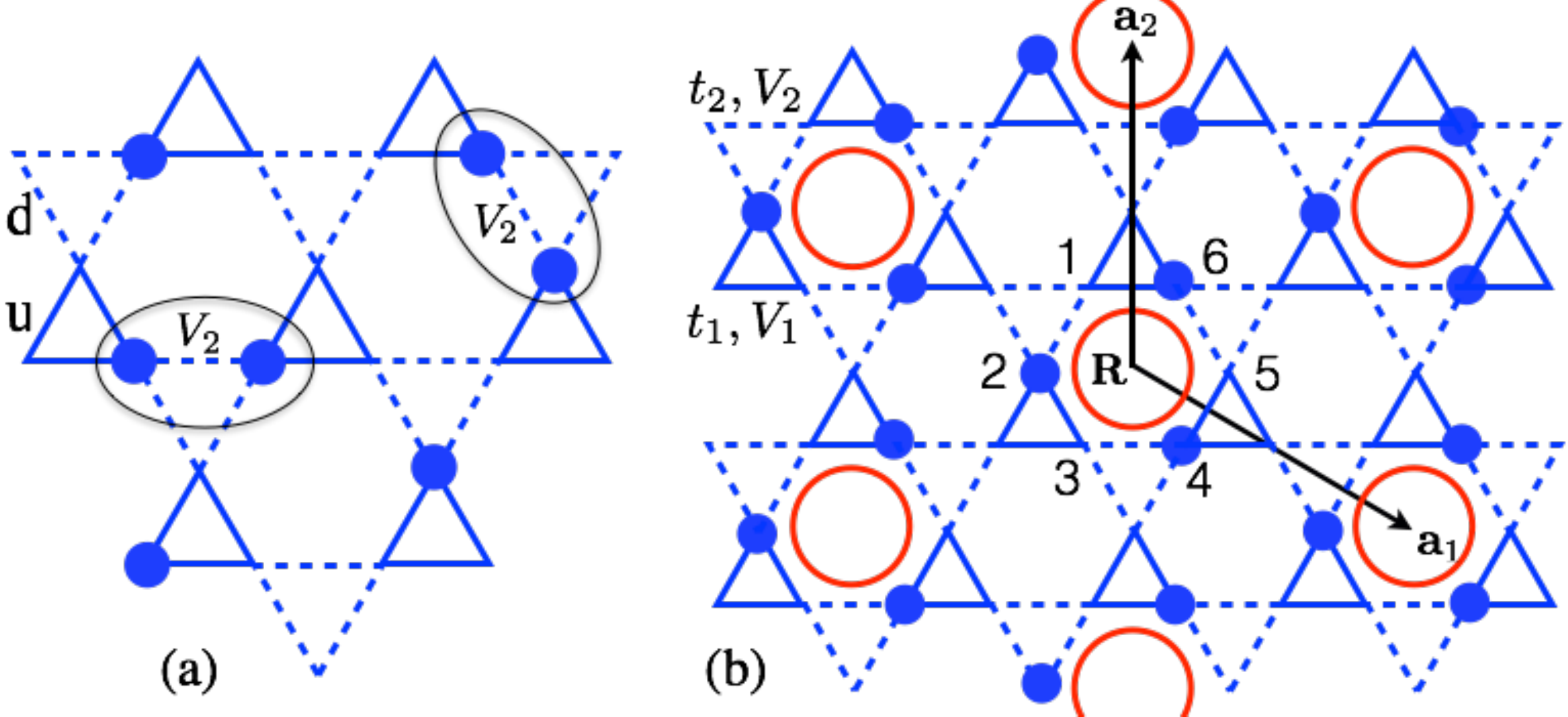}}
\caption{(Color online.) 
(a) The electron configuration in the CMI without the PCO 
when $V_2 \ll t_1$ and $V_1 \gg t_2$. 
(b) The electron configuration in the CMI with the PCO. 
Three electrons hop resonantly in each hexagon that is marked by a
(red) circle. These marked hexagons form an emergent triangular lattice
(with lattice vectors ${\bf a}_1, {\bf a}_2$). 
 }
\label{fig1}
\end{figure}

Recently LiZn$_2$Mo$_3$O$_8$ has been proposed as a new candidate 
material for QSL ground state~\cite{Sheckelton12,Sheckelton14,Mourigal14}. 
Besides the usual QSL phenomenology~\cite{Sheckelton12,Sheckelton14,Mourigal14},
the experiments reveal two Curie-Weiss regimes of the spin susceptibility 
at finite temperatures. The low temperature Curie-Weiss regime 
is governed by a much smaller Curie-Weiss temperature than the high temperature 
one and a reduced Curie constant which is 1/3 of the high temperature counterpart.  
In this work, to understand how one can achieve       
the QSL phenomenology and the puzzling two Curie-Weiss regimes in this material,
we consider a 1/6-filled extended Hubbard model with nearest-neighbor repulsions
on an anisotropic Kagome lattice. We first propose the existence of plaquette charge order
(PCO) for the charge degree of freedom. We emphasize that 
the emergence of the PCO reconstructs the spin state of the system. 
We further propose a U(1) QSL with spinon Fermi 
surfaces for the spin ground state and a PCO for the charge ground state 
in this system. The Mott insulators in partially filled 
systems arise due to the large nearest-neighbor repulsions~\cite{Chunhua09,Chen14} 
and localization of the charge degrees of freedom in certain cluster units. 
Hence such Mott insulators are dubbed ``cluster Mott insulators'' 
(CMIs)~\cite{Chen14,Lv2015,GangChen2014b}. 

The single-band extended Hubbard model is defined on the 
anisotropic Kagome lattice of the Mo sites (see Fig.~\ref{fig1})
and is given by~\cite{GangChen2014b}
\begin{eqnarray}
H &=&  \sum_{ \langle ij \rangle \in \text{u}} 
[-t_1 (c^{\dagger}_{i\sigma} c^{\phantom\dagger}_{j\sigma} + h.c.)  + V_1 n_i n_j  ] 
\nonumber \\
& + &  \sum_{ \langle ij \rangle \in \text{d}} 
[-t_2 (c^{\dagger}_{i\sigma} c^{\phantom\dagger}_{j\sigma} + h.c.)  + V_2 n_i n_j  ] 
\nonumber \\
&+&  \sum_{i} \frac{U}{2} (n_i -\frac{1}{2})^2 ,
\label{eq1}
\end{eqnarray}
where the spin index $\sigma$ ($ = \uparrow, \downarrow$) is implicitly summed, 
$c^{\dagger}_{i\sigma}$ ($c^{\phantom\dagger}_{i\sigma} $) creates 
(annihilates) an electron with spin $\sigma$ at lattice site $i$, and 
$t_1, V_1$ and $t_2, V_2$ are the nearest-neighbor electron hopping 
and interaction in the up-pointing triangles (denoted as `u') and 
the down-pointing triangles (denoted as `d'), respectively. 
$n_i = \sum_{\sigma} c^\dagger_{i\sigma}c^{\phantom\dagger}_{i\sigma}$ 
is the electron occupation number at site $i$. Since there exists only 
one unpaired electron in each Kagome lattice unit cell \cite{Sheckelton12}, 
the electron filling for this Hubbard model is $1/6$.

Although the down-triangles are larger in size than the up-triangles 
in LiZn$_2$Mo$_3$O$_8$, because of the large spatial extension
of the $4d$ Mo electron orbitals we think it is necessary to 
include the inter-site repulsion $V_2$ for the down-triangles.
For LiZn$_2$Mo$_3$O$_8$ we expect $t_1 > t_2$ and 
$U >V_1 \sim V_2$. Because of the very dilute electron filling, 
although the Hubbard $U$ is the largest energy scale, 
it alone can only remove double electron occupation on a single 
lattice site and {\it cannot} localize the electron. 
If there is no $V_1$ or $V_2$ interactions, 
the electrons can still transport on the lattice without encountering 
any electron double occupancy on a single lattice site. 
So we need $V_1$ and $V_2$ to localize the electrons in the (elementary) 
triangles of the Kagome lattice instead of the lattice sites.

Let us first explain the electron localization in the absence of $V_2$. 
Clearly, as $t_2$ is the hopping between the up-triangles, 
when $V_1\gg t_2$, the electrons are localized on the up-triangles
with one electron per up-triangle (see Fig.~\ref{fig1}a). 
In this picture, the localized electron can hop freely among 
the three lattice sites within each up-triangle and gain local 
kinetic energy $\sim \mathcal{O}(t_1)$ while the 
electron number on the down-triangle is strongly fluctuating.  
After $V_2$ is introduced, as $V_2$ increases, the configuration with more than one electrons 
on the down-triangles (like the one in Fig.~\ref{fig1}a) becomes less favorable energetically. 
When the interaction energy cost ($\sim \mathcal{O}(V_2)$) on the down-triangle 
overcomes the local kinetic energy gain $\sim \mathcal{O}(t_1)$,
the electron number on each down-triangle is also fixed to one
and the electrons can no longer move freely within each up-triangle. 
Instead, the electrons develop a collective motion. 
For instance, in Fig.~\ref{fig1}b, the three electrons on the hexagon
at position ${\bf R}$ can tunnel between the configuration occupying sites 1,3,5 
and the other configuration occupying sites 2,4,6. 
This collective electron tunnelling process 
preserves the electron number on each triangle and is the dominant
physical process below the Mott gap. 
We show this collective electron tunnelling gives rise to a 
long-range PCO that breaks the lattice symmetry spontaneously. 
With the PCO, the electrons are preferentially tunnelling back and forth 
on the hexagons that are marked with a (red) circle (see Fig.~\ref{fig1}b). 
We will refer these special hexagons as ``resonating'' hexagons. 
On these resonating hexagons, the three electrons form a linear superposition
state of the two electron configurations with sites 1,3,5 or 
sites 2,4,6 occupied. We emphasize and will show in Sec.~\ref{sec2} that
the emergence of the PCO in the CMI is a {\it quantum effect} and cannot
be obtained from the classical treatment of the electron interaction.  

With the PCO, 1/3 of the elementary hexagons become resonating.   
As shown in Fig.~\ref{fig1}b, these resonating hexagons form 
an emergent triangular lattice (ETL). The PCO triples the original 
unit cell of the Kagome lattice, and the localized electron number 
in the enlarged unit cell now becomes 3, which is odd. 
According to Hastings' theorem \cite{Hastings04}, 
the CMI with the PCO is not connected to a trivial band insulator 
and the QSL is expected. In the resulting U(1) QSL, we obtain 9 
mean-field spinon sub-bands, compared to the 3 spinon bands
in the U(1) QSL for the CMI without the PCO. 
The 9 mean-field spinon sub-bands are obtained by splitting
the 3 spinon bands of the CMI without the PCO, 
and this is the reason why we use the term ``{\it sub-bands}''. 
A direct band gap separates the lowest spinon sub-band 
from other spinon sub-band in the presence of the PCO.
The lowest spinon sub-band is completely filled by 2/3 of the spinons,
leaving the remaining 1/3 of the spinons to partially fill the second and third 
lowest spinon sub-band. Because of the band gap, the only active degrees 
of freedom at low energies are the spinons in the partially filled 
spinon sub-band, and the fully-filled 
lowest spinon sub-band is inert to external magnetic field at low temperatures 
as long as the PCO persists. Therefore, only 1/3 of the magnetic degrees of 
freedom are active at low temperatures. 
If one then considers the local moment formation starting 
from the band filling picture of the spinons 
(just like electrons occupying the same band structure) 
only the 1/3 of the spinons from the partially filled upper 
bands would participate in the local moment formation. This means 
the CMI with the PCO would be continuously connected to 
the Curie-Weiss regime with the 1/3 Curie constant 
at finite temperature (compared to the case when all 
spinons can participate in the local moment formation). 
This would explain the two Curie-Weiss regimes in the spin 
susceptibility data of LiZn$_2$Mo$_3$O$_8$. 
More precise connection to the existing 
experiments is discussed later. 

The rest of the paper is structured as follows.
In Sec.~\ref{sec2}, we show the CMI develops the PCO in the charge sector when
every triangle contains only one electron. 
We generalize the Levin-Wen variational string wavefunction approach \cite{Levin04} 
to study the reconstruction of the spinon band structure by the PCO in Sec.~\ref{sec3}. 
In Sec.~\ref{sec4}, we explain the consequence of this reconstructed spinon band structure  
and discuss the low-temperature magnetic susceptibility. 
In Sec.~\ref{sec5}, we connect our theory to the experiments
on LiZn$_2$Mo$_3$O$_8$, suggest possible future experiments,
and discuss other Mo based cluster magnets. 
Finally, we discuss the quantum chemistry justification of 
the extended Hubbard model in Appendix~\ref{asec1}.
A complementary explanation of the double Curie regimes based on the 
spin state reconstruction for 
the strong PCO regime is given in Appendix~\ref{asec2}. 
In Appendix~\ref{asec3}, we provide the detailed formalism of the 
mean-field theory in the main text. 
 
\section{The emergence of the plaquette charge order}
\label{sec2}

As a preparation step, we first employ the standard 
slave-rotor representation and map the low-energy charge 
sector Hamiltonian into a quantum dimer model on the dual honeycomb lattice.  
With the quantum dimer model, we then show the system 
should develop the PCO. 

Using standard slave-rotor representation \cite{Florens04,Lee05}, 
we first express the electron operator, 
$c^\dagger_{i\sigma} = f^\dagger_{i\sigma} e^{i \theta_i}$,
where the bosonic rotor ($e^{i\theta_i}$) carries the electron charge and
the fermionic spinon ($f^\dagger_{i\sigma}$) carries the spin quantum number.  
As it is well-known,
the slave-rotor representation enlarges the Hilbert space. 
To constrain the enlarged Hilbert space, we introduce an angular 
momentum variable $L_i^z$, $L_i^z = [\sum_{\sigma} f^\dagger_{i\sigma} 
f^{\phantom\dagger}_{i\sigma} ] - {1}/{2}$,
where $L_i^z$ is conjugate to the rotor variable with $[\theta_i,L_j^z] = i \delta_{ij}$. 
In terms of the slave-rotor variables, the extended Hubbard model is now reformulated as
\begin{widetext}
\begin{eqnarray}
H &=& \sum_{\langle ij \rangle \in \text{u} } 
\big[{-t_1} \big(e^{i (\theta_i - \theta_j ) } f^\dagger_{i\sigma} f^{\phantom\dagger}_{j\sigma} + h.c.\big)+ V_1 (L_i^z +\frac{1}{2})(L_j^z +\frac{1}{2})\big] 
\nonumber \\
& + & \sum_{\langle ij \rangle \in \text{d} } 
\big[{-t_2} \big(e^{i (\theta_i - \theta_j ) } f^\dagger_{i\sigma} f^{\phantom\dagger}_{j\sigma} + h.c.\big)+ V_2 (L_i^z +\frac{1}{2})(L_j^z +\frac{1}{2})\big] 
\nonumber \\
& + & \sum_i \big[\frac{U}{2} (L_i^z)^2  + h_i (L_i^z +\frac{1}{2} - \sum_{\sigma} f^\dagger_{i\sigma} 
f^{\phantom\dagger}_{i\sigma})\big] , 
\end{eqnarray}
where we have introduced $h_i$ as a Lagrange multiplier to impose
the Hilbert space constraint. since the on-site interaction $U$ is assumed to be the largest energy scale, in the large $U$ limit the double electron occupation is always suppressed. 
Hence, the angular momentum variable $L_i^z$ primarily takes 
$ L_i^z = 1/2$ ($-1/2$) for a singly-occupied (empty) site. 

Via a decoupling of the electron hopping terms in $H$ into the spinon and rotor sectors, 
we obtain the following two coupled Hamiltonians for the spin and charge sectors, respectively,
\begin{eqnarray}
H_{\text{s}} &=& 
-\sum_{\langle ij \rangle} t^{\text{eff}}_{ij}
 (f^{\dagger}_{i\sigma} f^{\phantom\dagger}_{j\sigma} + h.c.)
 - \sum_{i} h_i  
 f^\dagger_{i\sigma} f^{\phantom\dagger}_{i\sigma} ,
 \label{eq5}
\\
H_{\text{c}} &=&  \sum_{\langle ij \rangle } 
\big[ {-2}{J}^{\text{eff}}_{ij} \cos (\theta_i -\theta_j) 
+ V_{ij} (L_i^z +\frac{1}{2})
  (L_j^z +\frac{1}{2}) \big]
+ \sum_i \big[\frac{U}{2} (L_i^z)^2  + h_i (L_i^z +\frac{1}{2})\big] , 
\label{eq6}
\end{eqnarray}
where $t_{ij}^{\text{eff}} = t_{ij} \langle e^{i \theta_i - i \theta_j} \rangle 
\equiv |t_{ij}^{\text{eff}}| e^{i a_{ij}}$,
${J}_{ij}^{\text{eff}} = t_{ij} \sum_{\sigma} 
\langle f^\dagger_{i\sigma} f^{\phantom\dagger}_{j\sigma} \rangle 
\equiv   |  { J}_{ij}^{\text{eff}}  | e^{- i a_{ij}}$
and $t_{ij} = t_1$ ($t_2$), $V_{ij} = V_1$ ($V_2$) 
for the bond $ij$ on the up-triangles (down-triangles).
 Here, we have chosen the mean-field ansatz to 
respect the symmetries of the Kagome lattice. 
The Hamiltonians $H_{\text{s}}$ and $H_{\text{c}}$ 
are invariant under an internal U(1) gauge transformation, 
$f^{\dagger}_{i\sigma} \rightarrow f^{\dagger}_{i\sigma} e^{- i \chi_i},
\theta_i \rightarrow \theta_i + \chi_i$, and 
$a_{ij} \rightarrow a_{ij} + \chi_i - \chi_j$.

\begin{figure}[t]
{\includegraphics[width=6cm]{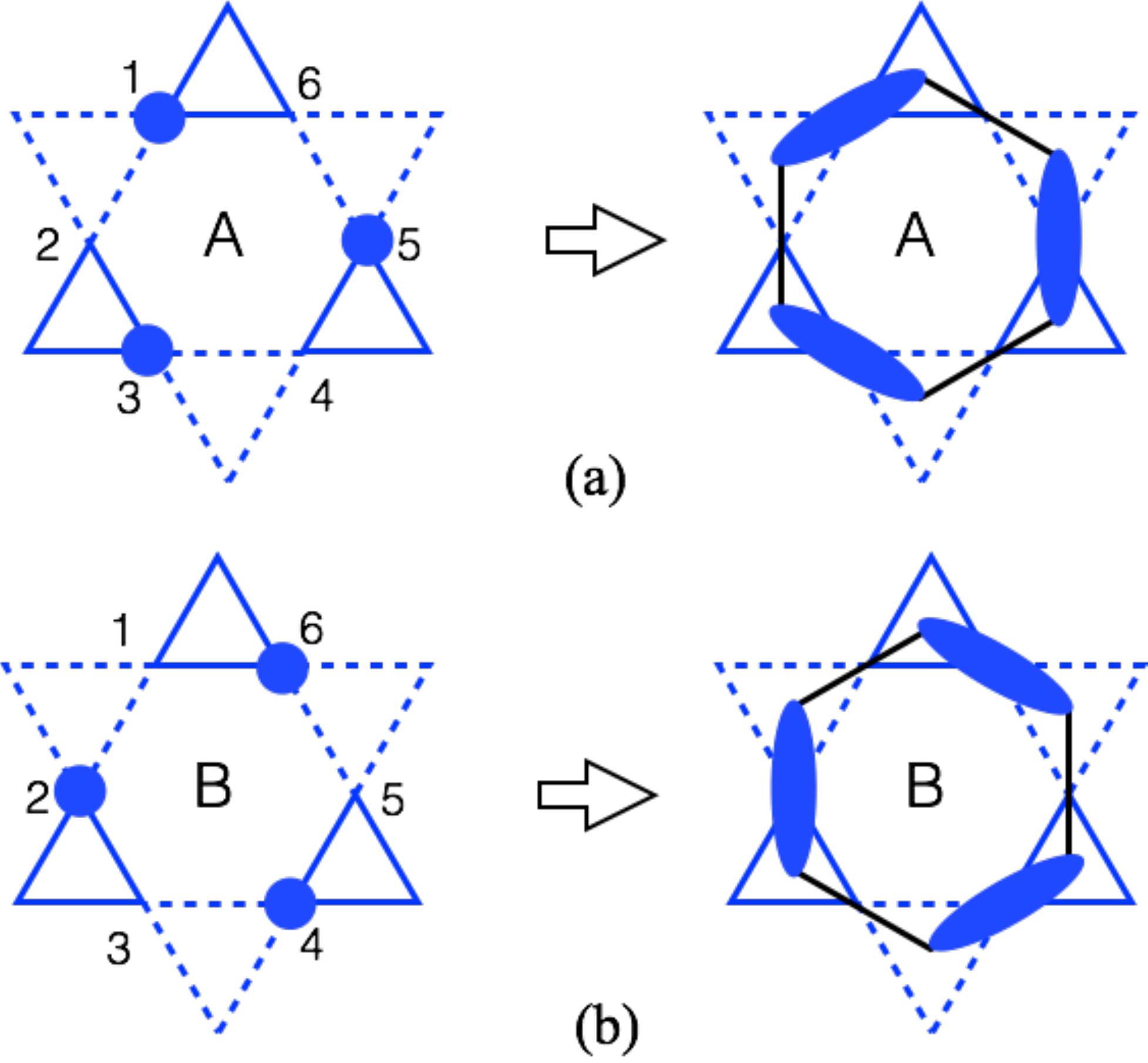}}
\caption{(Color online.) 
The two collective hopping processes that contribute to the 
ring electron hopping or the ring exchange in Eq.~\eqref{Hring}. 
The (blue) solid ball represents the electron or the charge rotor. 
}
\label{fig2}
\end{figure}

We now focus on the charge sector 
and study the low energy physics of the charge sector. 
From the previous slave-rotor formulation, 
the charge sector Hamiltonian is given by 
\begin{eqnarray}
H_{\text{c}} & =&  
\sum_{\langle ij\rangle } 
{- 2J_{ ij}^{\text{eff}}}\cos (\theta_i -\theta_j) 
+ V_{ ij}( L_i^z + \frac{1}{2})  (L_j^z + \frac{1}{2}) 
+ \sum_i h_i (L_i^z + \frac{1}{2}),
\end{eqnarray}
where we have dropped the $U$ interaction term with the understanding that  
$L_i = \pm 1/2$ in the large $U$ limit.  
This charge sector Hamiltonian can be thought as a Kagome lattice 
spin-1/2 XXZ model in the presence of an external magnetic field upon
identifying the rotor operators as the effective spin ladder operators,
$e^{\pm i \theta_i } = L^{\pm}_i $
where $ L^{\pm}_i |  L_i^z = \mp \frac{1}{2} \rangle 
= | L_i^z =  \pm \frac{1}{2} \rangle$. 
Thus the corresponding effective spin-$L$ model reads
\begin{eqnarray}
H_{\text{c}} & = &
 \sum_{ \langle ij \rangle   } \big[ 
{ - J_{ ij}^{\text{eff}}}(  L^+_i L^-_j + h.c. ) 
  + V_{ ij} L_i^z L_j^z \big]
 + 
\tilde{h} \sum_i L_i^z, 
\end{eqnarray}
in which we have made a uniform mean-field approximation such that 
$h_i + 3(V_1+V_2) \equiv \tilde{h} $. 
The 1/6 electron filling is equivalent to 
${N_s}^{-1} \sum_i L_i^z = -  {1}/{6}$, 
where $N_s$ is the total number of Kagome lattice sites. 
\end{widetext}

\begin{figure}[t]
{\includegraphics[width=6cm]{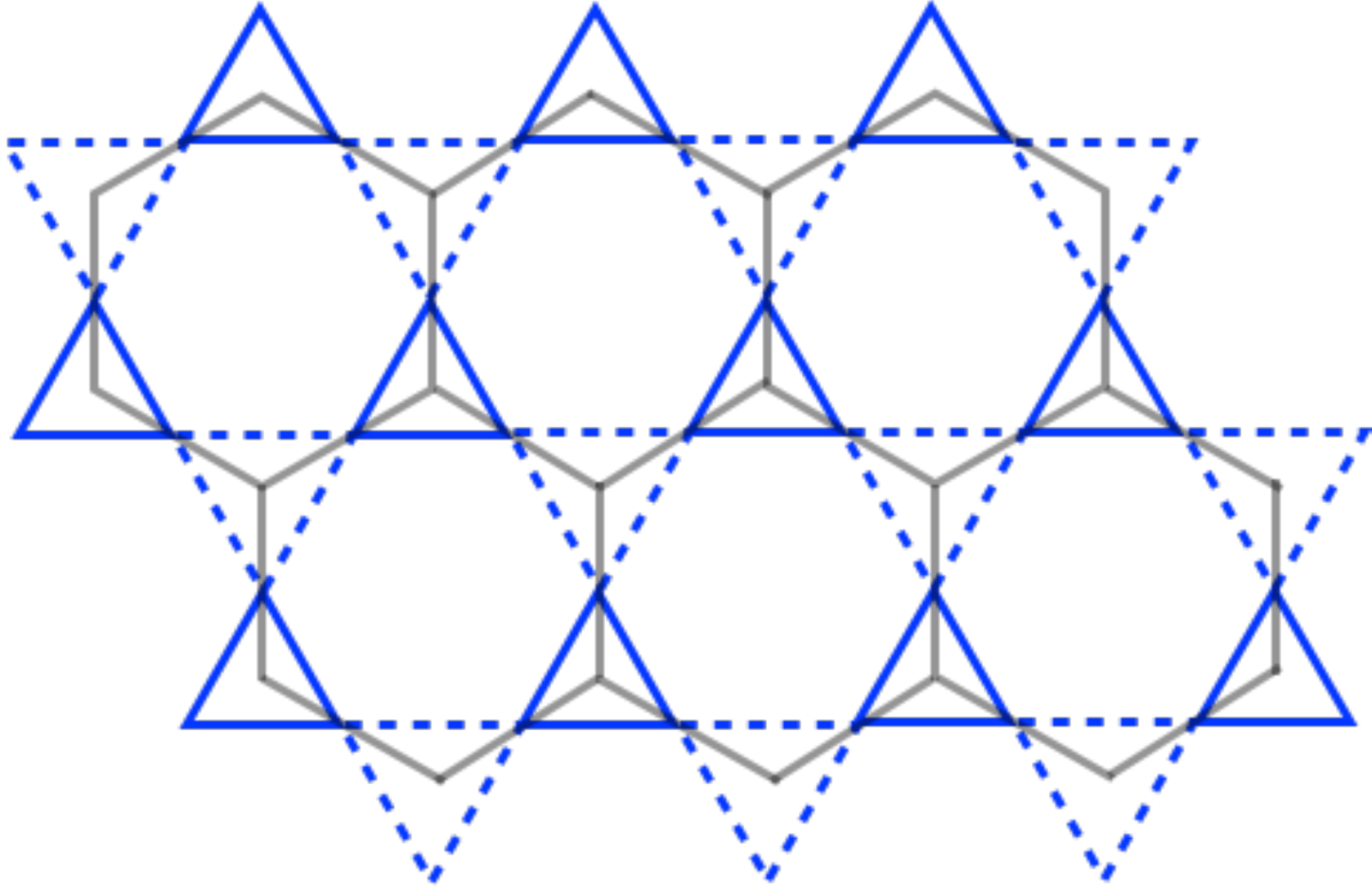}}
\caption{(Color online.) 
The anisotropic Kagome lattice and the dual honeycomb lattice (in gray).}
\label{fig3}
\end{figure}

As we explained in Sec.~\ref{sec1}, when $V_1$ ($V_2$) is large enough compared
to $t_2$ ($t_1$), the electron number on each triangle, both up-triangle and down-triangle, is fixed to be one. The electron occupation configuration that satisfies this condition 
is highly degenerate. The presence of the electron hopping, i.e. $L^+_i L^-_j$, lifts
this classical degeneracy and the effective interaction can be obtained from a 
third-order degenerate perturbation theory. The resulting effective ring exchange 
Hamiltonian is given as 
\begin{equation}
H_{\text{c,ring}} = - \sum_{\hexagon } J_{\text{ring}} 
 (L^+_1 L^-_2 L^+_3 L^-_4 L^+_5 L^-_6 + h.c. ),
\label{Hring}
\end{equation}
where ``$\hexagon$'' refers to the elementary hexagon of the Kagome lattice,  
$J_{\text{ring}} = 
{6 (J_1^{\text{eff}})^3 }/{ V_2^2 } +  {6 (J_2^{\text{eff}} )^3 }/{ V_1^2}
$ and ``1, $\cdots$, 6'' are the 6 vertices on the perimeter of the elementary hexagon 
(see Fig.~\ref{fig2}). 
This ring Hamiltonian in Eq.~\ref{Hring} 
describes the collective tunnelling of three electron charges
between A and B configurations in Fig.~\ref{fig2}. 

We now map $H_{\text{c,ring}}$ into a quantum dimer model on the dual honeycomb 
lattice that is formed by the centers of the triangles on the Kagome lattice (see Fig.~\ref{fig3}).  
As depicted in Fig.~\ref{fig2}, a dimer is placed on the corresponding 
link if the center of the link (or the Kagome lattice site) is occupied 
by an electron charge. The rotor operator $L_i^{\pm}$ simply adds or removes 
the charge dimer. So $H_{\text{c,ring}}$ is mapped into the quantum dimer model
with only a resonant term,
\begin{equation}
H_{\text{c,ring}} = - J_{\text{ring}} 
 \sum_{ \varhexagon }
( | \varhexagon_{\text A}\rangle \langle \varhexagon_{\text B} | + 
| \varhexagon_{\text B}\rangle \langle \varhexagon_{\text A} | )
\label{eq64}
\end{equation}
where $| \varhexagon_{\text A} \rangle$ and $| \varhexagon_{\text B} \rangle$
refer to the two charge dimer covering configurations in the elementary hexagon 
``$\varhexagon$'' of the dual honeycomb lattice as shown in Fig.~\ref{fig3}. 

\begin{figure}[t]
{\includegraphics[width=7cm]{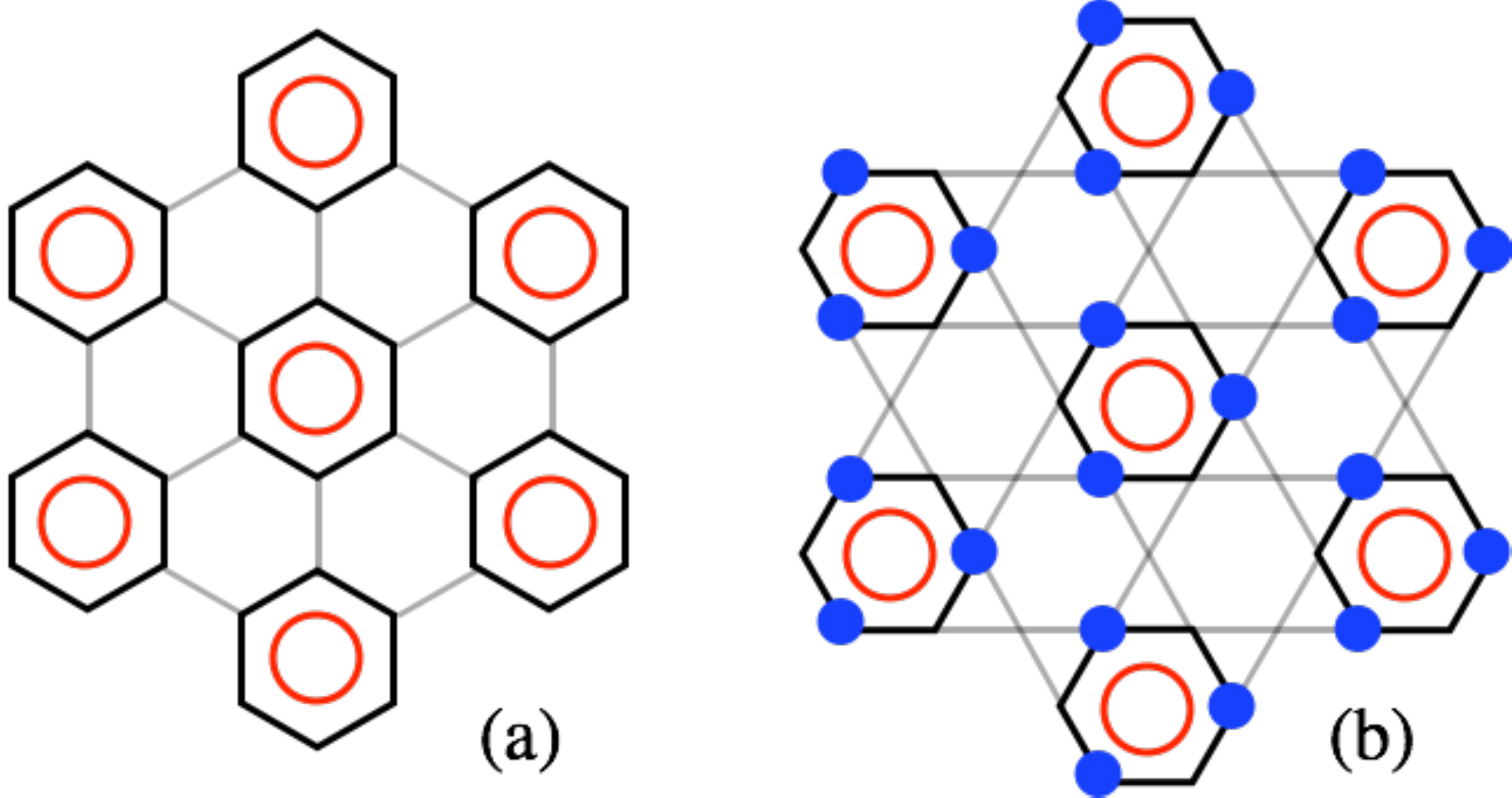}}
\caption{(Color online)
(a) The plaquette charge dimer ordering pattern on the dual honeycomb lattice. 
The charge dimers have a high probability to occupy the bold bonds of the 
resonating hexagons (see the main context). 
(b) The corresponding PCO on the Kagome lattice. 
We mark the resonating hexagons with both dark bonds and the red circles.   
}
\label{fig4}
\end{figure}

In Ref.~\onlinecite{Moessner01}, Moessner, Sondhi and Chandra studied 
the phase diagram of the quantum dimer model on the honeycomb lattice 
quite extensively. In the case with only the resonant term  
of our model in Eq.~\eqref{eq64}, they found a 
translational symmetry breaking phase with a plaquette dimer order,
in which the system preferentially gains dimer resonating (or kinetic) energy 
through the resonating hexagons on the dual honeycomb lattice (see Fig.~\ref{fig4}a).
The dimers on resonating hexagons form a linear superposition of the dimer covering
configurations $|\varhexagon_{\text A} \rangle$ and $|\varhexagon_{\text B} \rangle$.
This indicates that our model is unstable to translational symmetry breaking. 
The plaquette dimer order of the quantum dimer model is then mapped back to 
the plaquette charge order (PCO) on the Kagome lattice (see Fig.~\ref{fig4}b).
Just like the resonating dimers, the three electron charges 
on the resonating hexagons also form a linear superposition 
of two occupation configurations 
in Fig~\ref{fig4}b. This is a quantum mechanical effect and 
cannot be obtained by treating the inter-site electron 
interactions $V_1$ and $V_2$ in a classical fashion. 
Moreover, this PCO can be regarded as a {\it local charge resonating valence bond} 
(RVB) state which contrasts with the spin singlet RVB of Anderson \cite{Anderson73,ANDERSON06031987}.

We note that similar type of PCO has already been obtained for extended Hubbard models with 
fermions or hard-core bosons on an {\sl isotropic} Kagome lattice with 1/3 and 2/3 fillings
in certain parameter regimes in previous works~\cite{Pollmann08,Fiete11,Pollmann14,Ferhat14}.
The result was obtained either through perturbatively mapping to the 
quantum dimer model or by a Hatree-Fock mean-field calculation. 
In particular, Ref.~\onlinecite{Isakov08} applied the quantum Monte Carlo technique 
to simulate a hardcore boson Hubbard model on an isotropic Kagome lattice and
discovered the PCO for 1/3 and 2/3 boson fillings. 
Because our model is defined on the anisotropic kagome lattice, it is not exactly the same
as the previous works.

With the PCO, the electrons are preferentially hopping around the 
perimeters of the resonating hexagons on the Kagome lattice. 
These resonating hexagons are periodically arranged, forming an emergent
triangular lattice (see Fig.~\ref{fig1}b). Due to the translational symmetry breaking, 
this emergent triangular lattice (ETL) has an enlarged unit cell that includes 9 sublattices
compared to 3 sublattices in a Kagome lattice (see Fig.~\ref{fig5}).

\section{The spinon band structure}
\label{sec3}

In Sec.~\ref{sec2}, using a slave-rotor approximation, we have shown that 
the system is unstable to the development of the PCO in the CMI where both 
up-triangle and down-triangle contain only one electron. 
Although our result is obtained by first starting from 
a translationally invariant mean-field 
ansatz, as we argue below, the PCO breaks 
the translational symmetry of the spinon mean-field state and the 
modified spinon band structure makes the PCO even more stable \footnote{We do not consider the possibility of the ferromagnetic ordering
in the extreme limit $V_1 = V_2 \gg t_1 = t_2$ and 
$U\rightarrow \infty$ that is considered in Ref.~\onlinecite{Pollmann08}
because this FM state is very unstable to the introduction
of the antiferromagnetic  
spin interaction between the electron spins }.

\vspace{0.4cm}

\subsection{Spin charge coupling in the CMI with the PCO}
\label{sec3A}

To understand how the PCO in the charge sector influences the spinon sector,
we first consider the low-energy effective ring hopping model 
in the CMI where both up-triangle and down-triangle contain only one electron,
\begin{widetext}
\begin{eqnarray}
H_{\text{ring}} &=& 
- \sum_{\hexagon} \sum_{\alpha \beta \gamma }
\big[
{\mathbb K}_1  (c^{\dagger}_{1\alpha} c^{\phantom\dagger}_{6\alpha} 
                                 c^{\dagger}_{5\beta} c^{\phantom\dagger}_{4\beta}  
  c^{\dagger}_{3\gamma} c^{\phantom\dagger}_{2\gamma}    + h.c.) 
+
{\mathbb K}_2  (c^{\dagger}_{1\alpha} c^{\phantom\dagger}_{2\alpha} 
                                 c^{\dagger}_{3\beta} c^{\phantom\dagger}_{4\beta}  
                                  c^{\dagger}_{5\gamma} c^{\phantom\dagger}_{6\gamma}  
+ h.c.) 
\big],
\label{eq65}
\end{eqnarray}
where $ {\mathbb K}_1 = 6t_1^3/V_2^2$ and ${\mathbb K}_2 = 6t_2^3/V_1^2 $
are readily obtained from the third-order degenerate perturbation theory. 
Here, $\alpha,\beta,\gamma = \uparrow,\downarrow$,
and ``1,$\cdots$,6'' are the 6 vertices in the elementary hexagon of the Kagome lattice.

Using the slave-rotor representation in Sec.~\ref{sec2} 
for the electron operator $c^\dagger_{i\alpha} =f_{i\alpha}^\dagger e^{i \theta_i} 
\equiv  f_{i\alpha}^\dagger L_i^+ $, the 
ring hopping model $H_{\text{ring}}$ can be decoupled as 
\begin{eqnarray}
\bar{H}_{\text{ring}}&=&
-  \sum_{\hexagon} \big[ \mathbb{K}_1
( L_1^+ L_2^- L_3^+ L_4^- L_5^+ L_6^ - \times
M_{165432} 
  + h.c.) + \mathbb{K}_2 
( L_1^+ L_2^- L_3^+ L_4^- L_5^+ L_6^ - \times
 M_{123456}  + h.c.) \big]
  \label{eq66a}
  \\
 & \equiv &  
 -\sum_{\varhexagon} \big[ \mathbb{K}_1 (
 |{ \varhexagon}_{\text A}\rangle  \langle { \varhexagon}_{\text B} | M_{165432} + 
|{ \varhexagon}_{\text B}\rangle  \langle { \varhexagon}_{\text A} | 
M_{165432}^{\ast}
)
 + \mathbb{K}_2 (
 |{ \varhexagon}_{\text A}\rangle  \langle { \varhexagon}_{\text B} | M_{123456} +
|{ \varhexagon}_{\text B}\rangle  \langle { \varhexagon}_{\text A} | 
 M_{123456} ^{\ast}
 ) \big], 
 \label{eq66b}
\end{eqnarray}
where 
$|\varhexagon_{\text A} \rangle$ and $|\varhexagon_{\text B}\rangle$ 
are the two charge dimer coverings in Fig.~\ref{fig2}. 
Here we are focusing on the charge sector and treating the 
spinon sector in a mean-field fashion, i.e.
\begin{equation}
M_{ijklmn}  =
\sum_{\alpha\beta\gamma}
\langle f^{\dagger}_{i\alpha} f^{\phantom\dagger}_{j\alpha} 
  f^{\dagger}_{k\beta} f^{\phantom\dagger}_{l\beta}  
  f^{\dagger}_{m\gamma} f^{\phantom\dagger}_{n\gamma}  \rangle,
\end{equation}
where the lattice sites $i,j,k,l,m,n$ are arranged either clockwise 
or anti-clockwise and $ M_{ijklmn}^{\ast}$ is the complex conjugate of $M_{ijklmn}$. 
What we did in Eq.~\eqref{eq66b} is to directly 
couple the charge sector quantum dimer model
with the spinon sector. By doing this, we can study how the 
charge sector is influenced by the spinon sector.

\begin{figure}[t]
{\includegraphics[width=10cm]{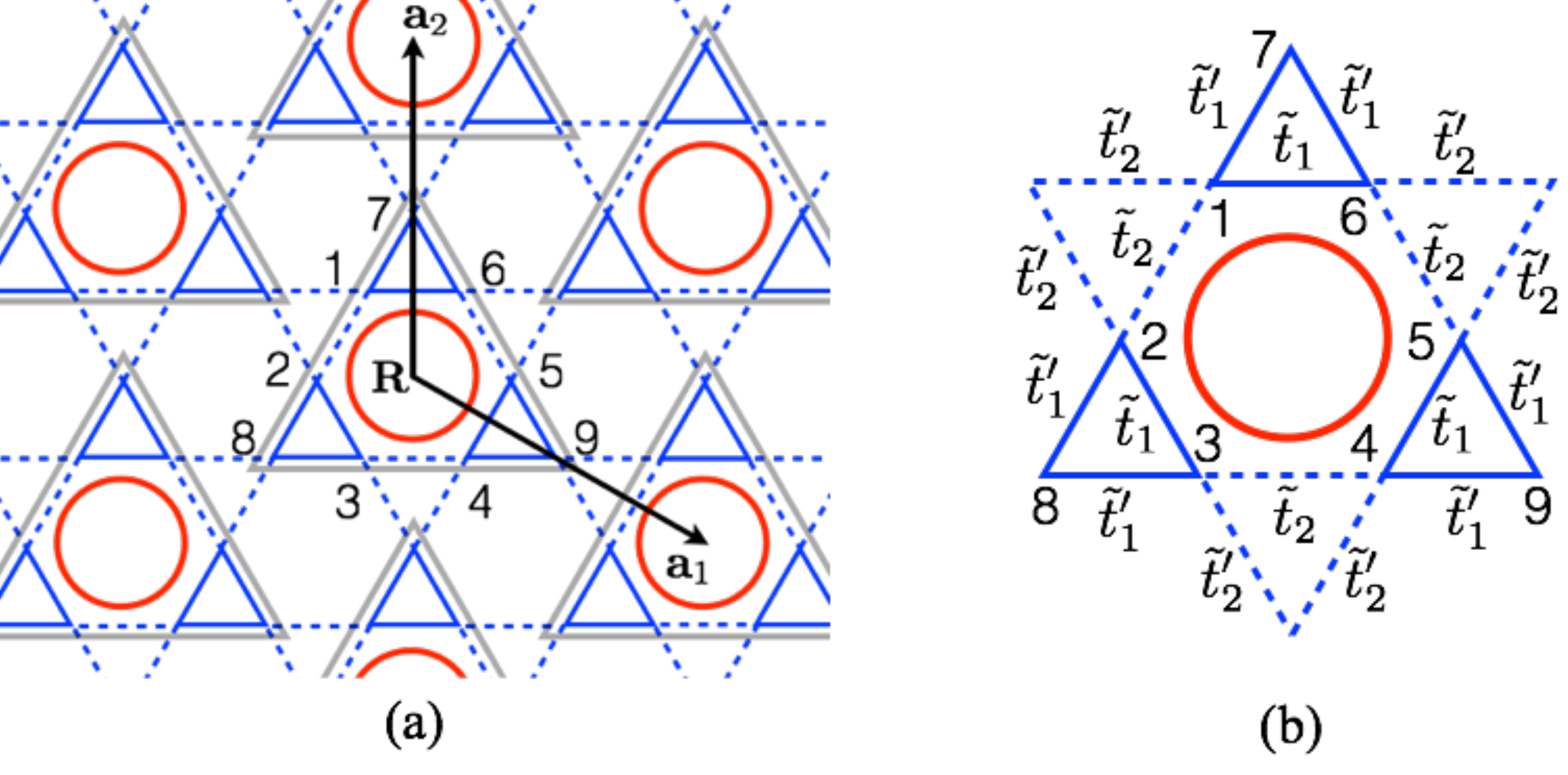}}
\caption{(Color online) (a)
The Kagome lattice is partitioned into unit cells on the ETL.  
The unit cell (marked by a gray triangle) contains 9 sublattices
that are labelled by ``1,2,$\cdots$,8,9''. 
(b) The spinon hoppings on the bonds that surround 
around a resonating hexagon. }
\label{fig5}
\end{figure}

The parameter $M_{ijklmn}$ is evaluated in the spinon mean-field ground state, which 
we explain below. For a time-reversal invariant system, we expect
$M_{ijklmn} = M_{ijklmn}^{\ast}$. 
If we assume the spinon sector respects the translational invariance of the Kagome lattice, 
the resulting charge sector model would be equivalent to $H_{\text{c,ring}}$
in Eq.~\eqref{Hring} and also to the quantum dimer model in Eq.~\eqref{eq64}
except for the renormalized couplings, and this would immediately imply the system
should develop the PCO and spontaneously break the translational symmetry of the Kagome lattice. In turn, the breaking of lattice symmetry by the PCO would then  influence the spinon sector. To understand
how the spinon band structure is modified by the underlying PCO 
in the charge sector and how the modified spinon band structure 
feeds back to the charge sector, we take the enlarged unit cell of the
ETL and introduce the following spinon mean-field  
(hopping) Hamiltonian (see Fig.~\ref{fig5}b),
\begin{eqnarray}
 \bar{H}_{\text{s}}  
  &=&
 \sum_{{\bf R}} \big[
 { -\bar{t}_1 } \big( 
  f^\dagger_{{\bf R}1\sigma }  f^{\phantom\dagger}_{{\bf R}6\sigma }  
+     f^\dagger_{{\bf R} 2\sigma } f^{\phantom\dagger}_{{\bf R}3\sigma }  
+    f^\dagger_{{\bf R} 4\sigma }  f^{\phantom\dagger}_{{\bf R}5\sigma }  
     \big)
-
     \bar{t}_2 \big( f^\dagger_{{\bf R}1\sigma }   f^{\phantom\dagger}_{{\bf R}2\sigma } 
     +  f^\dagger_{{\bf R}3\sigma }  f^{\phantom\dagger}_{{\bf R}4\sigma }  
     +  f^\dagger_{{\bf R}5\sigma }  f^{\phantom\dagger}_{{\bf R}6\sigma }   
     \big)
     \nonumber \\
&-& \bar{t}_1' \big(
         f^\dagger_{{\bf R}1\sigma }   f^{\phantom\dagger}_{{\bf R}7\sigma }  
   +       f^\dagger_{{\bf R}6\sigma }  f^{\phantom\dagger}_{{\bf R}7\sigma }  
   +        f^\dagger_{{\bf R}2\sigma }  f^{\phantom\dagger}_{{\bf R}8\sigma }  
     +       f^\dagger_{{\bf R}3\sigma }  f^{\phantom\dagger}_{{\bf R}8\sigma } 
+
         f^\dagger_{{\bf R}9\sigma }  f^{\phantom\dagger}_{{\bf R}4\sigma }  
   +      f^\dagger_{{\bf R}9\sigma } f^{\phantom\dagger}_{{\bf R}5\sigma }   
      \big) -\bar{t}_2' \big(
            f^\dagger_{{\bf R} 9\sigma }  f^{\phantom\dagger}_{{\bf R} + {\bf a}_1,1\sigma }
      \nonumber \\
  &+&
   f^\dagger_{{\bf R}9\sigma }   f^{\phantom\dagger}_{ {\bf R}+ {\bf a}_1,2\sigma } 
 +  f^\dagger_{{\bf R}7\sigma } f^{\phantom\dagger}_{{\bf R} + {\bf a}_2,3\sigma } 
  +   f^\dagger_{{\bf R}7\sigma }  f^{\phantom\dagger}_{{\bf R} + {\bf a}_2,4\sigma }
+    f^\dagger_{{\bf R}8\sigma }  f^{\phantom\dagger}_{{\bf R} -{\bf a}_1-{\bf a}_2,5\sigma }      
         +   f^\dagger_{{\bf R}8\sigma } f^{\phantom\dagger}_{{\bf R} -{\bf a}_1-{\bf a}_2,6\sigma }   
    \big) + h.c. \big], 
\label{eq68}
\end{eqnarray}
where ${\bf R}$ labels the unit cell of the ETL and
``1, 2, $\cdots$, 8, 9'' label the 9 sublattices of the ETL. 
This choice of spinon hopping parameters respects the 3-fold rotation symmetry and 
the reflection symmetry of the resonating hexagons (see Fig.~\ref{fig5}b).  
Moreover, in Eq.~\eqref{eq68}, the spinon hoppings are related to 
the charge sector via
\begin{eqnarray}
\bar{t}_1 &=& t_1 \langle L^+_1 ({\bf R}) L^-_6 ({\bf R})   \rangle 
\label{eq27a}
\\
\bar{t}_2 &=& t_2 \langle L^+_1 ({\bf R}) L^-_2 ({\bf R})   \rangle 
\label{eq27b}
\\
\bar{t}_1' &=& t_1  \langle L^+_1 ({\bf R}) L^-_7({\bf R})   \rangle 
\label{eq27c}
\\
\bar{t}_2' &=& t_2  \langle L^+_9 ({\bf R}) L^-_1({\bf R} + {\bf a}_1)   \rangle 
\label{eq27d}
\end{eqnarray}
such that the influence of the charge sector on the spinon sector is captured. 

 \end{widetext}

In the presence of the PCO, we expect $\bar{t}_1 > \bar{t}_1'$
and $\bar{t}_2 > \bar{t}_2'$ due to the presence of the PCO. 
According to Eqs.~\eqref{eq27a},~\eqref{eq27b}, \eqref{eq27c} and 
\eqref{eq27d}, the presence of the PCO would enhance the bonding of the charge rotors and then the spinon hoppings 
in the resonating hexagons and weakens the ones in the non-resonating hexagons. 
The enhanced spinon hoppings in the resonating hexagons 
further strengthen the couplings of the $\bar{H}_{\text{ring}}$
in the resonating hexagons through $M_{ijklmn}$ in Eq.~\eqref{eq66b}.
Thus the PCO would become more stable if the coupling between
spinon and charge excitations is switched on.

\subsection{Generalized Levin-Wen's variational dimer wavefunction approach and the spinon band structure in the presence of the PCO}
\label{sec3B} 

We now consider the combination of the spinon hopping model $\bar{H}_{\text{s}}$ 
in Eq.~\eqref{eq68} with the ring hopping model $\bar{H}_{\text{ring}}$ in Eq.~\eqref{eq66b}. 
It was pointed by Levin and Wen \cite{Levin05,Levin04} that 
quantum dimer model is an example of string-net models. 
In Ref.~\onlinecite{Levin04}, Levin and Wen developed a  
variational string wavefunction approach (or string mean-field theory)
to describe the properties of quantum dimer model. 
To make the nomenclature consistent,  
we refer Levin-Wen's variational string wavefunction approach
as variational dimer wavefunction approach in the following. 
Since the charge sector is described by a quantum dimer model, 
we can extend Levin-Wen's variational string wavefunction approach \cite{Levin04} 
to solve the coupled charge and spinon problem in Sec~\ref{sec3A}. 
In Levin and Wen's original work, the
variational dimer wavefunction approach was designed for 
{\it pure} quantum dimer model. 
The new ingredients of our problem are the presence of 
the spinon degrees of freedom, and the coupling and the mutual feedback 
between the spinons and charge dimers. 

We describe below 
the variational dimer wavefunction approach that is used to optimize the 
Hamiltonian $\bar{H}_{\text{ring}}$ in Eq.~\eqref{eq66b}
for the charge dimers. 
Following Levin and Wen, the variational dimer wavefunction is 
parametrized by a set of 
variational parameters $\{ z_i \}$ where $z_i$
is defined on each link 
of the dual honeycomb lattice. 
Here the links on the dual honeycomb lattice are also parametrized by the 
Kagome lattice sites that are located at the centers of the links. 
These variational parameters $z_i$ are also termed as string
(or dimer) fugacity by Levin and Wen \cite{Levin04}. 
For each set of $ \{ z_i \}$, the variational dimer wavefunction is given by 
\begin{equation}
\Psi ( \{ z_i \} ) = \prod_i \frac{|0 \rangle_i  + z_i | 1 \rangle_i}{ (1+ |z_i |^2)^{\frac{1}{2}}},
\label{eqfct}
\end{equation}
where $|0 \rangle_i$ and $| 1 \rangle_i$ define the 
absence and presence of the electron
charge at the Kagome lattice site $i$ or the dimer 
on the corresponding link on the dual honeycomb lattice, respectively. 
Moreover, we have the following relations by definition,
\begin{eqnarray}
&& n_i | 0 \rangle_i = 0,
\quad\quad\,\,   n_i | 1 \rangle_i = | 1 \rangle_i ,
\\
&& L_i^+ |0 \rangle_i = |1 \rangle_i ,
\quad  L_i^- |1 \rangle_i = |0 \rangle_i ,
\end{eqnarray}
where $n_i$ counts the electron number (or the number of dimers) at site $i$. 

We employ the symmetry of the PCO to reduce the number of free 
variational parameters in the dimer wavefunction $\Psi ( \{ z_i \} ) $. 
Using the symmetries of the ETL, we find that
only two variational parameters are needed
\begin{eqnarray}
&& z_{1} ({\bf R}) = z_{2} ({\bf R})  = z_{3}({\bf R})  = z_{4}({\bf R})
\nonumber  
\\ &&\quad\quad\quad   = z_{5} ({\bf R})  = z_{6} ({\bf R})  \equiv z,
\\
&& z_{7}({\bf R}) = z_{ 8}({\bf R}) = z_{ 9}({\bf R}) \equiv \tilde{z}, 
\end{eqnarray}
where 
$z_{\mu} ({\bf R})$ ($\mu = 1,2,\cdots, 9$) refers to the variational parameter of the 
$\mu$th sublattice at the unit cell ${\bf R}$ (see Fig.~\ref{fig5}). 
We have reduced the set of variational parameters in the
variational dimer wavefunction 
to $z$ and $\tilde{z}$. 
Moreover, $z$ and $\tilde{z}$ are not independent from each other. 
This is because of the charge localization constraint, {\it i.e.}
{\it every triangle contains only one electron}. 
In terms of the dimer language, this constraint is that every dual honeycomb lattice
site is connected by only one dimer. To satisfy this constraint, 
we only need to require
\begin{equation}
\langle n_1( {\bf R}) \rangle + \langle n_6( {\bf R}) \rangle + 
\langle n_7( {\bf R}) \rangle =1,
\end{equation}
where the expectation value is taken for the variational 
wavefunction $\Psi ( \{ z_i \})$. This relation connects $\tilde{z}$ to $z$. 

For the quantum dimer model $H_{\text{c,ring}}$ in Eq.~\eqref{eq64}, 
variational (or mean-field) phase is obtained by evaluating the Hamiltonian
$H_{\text{c,ring}}$ with respect to $\Psi(\{z_i\})$ and optimizing
the energy by varying $z$. 
This static variational approach, however, cannot directly produce the 
plaquette ordered phase of the quantum dimer model.  
What it gives is a translationally invariant variational ground state. 
To obtain the right result, Levin and Wen developed a dynamical 
variational approach. Namely, for each {\it static} variational ground state,
one checks the stability of the variational phase by considering the 
quantum fluctuation of this phase.
In the model that Levin and Wen were considering \cite{Levin04}, 
they found some modes in a translationally invariant variational ground state 
can become unstable and 
drive a dimer crystal ordering. We expect similar physics should happen to 
our quantum dimer model $H_{\text{c,ring}}$. 

Unfortunately, the dynamical variational approach by Levin and Wen 
is not a self-consistent variational approach and cannot be extended 
to the combined spinon and charge dimer problem that we are interested in here. 
Since we know our quantum dimer model $H_{\text{c,ring}}$ gives 
the ground state with the PCO
and we have argued that coupling the charge (dimer) with the spinons
makes the PCO even more stable in Sec.~\ref{sec3A},  
we now introduce the PCO into the system by explicitly breaking the lattice symmetry.
That is, we modify the ring hoppings ${\mathbb K}_1$ 
and ${\mathbb K}_2$ in $\bar{H}_{\text{ring}}$ of Eq.~\eqref{eq66a}.  
For the resonating hexagons, we change 
\begin{equation}
\mathbb{K}_1 \rightarrow \mathbb{K}_1 (1+ \delta),
\quad  \mathbb{K}_2 \rightarrow \mathbb{K}_2 (1 + \delta) ,
\end{equation}
and for the non-resonating hexagons, we use
\begin{eqnarray}
&& \mathbb{K}_1 \rightarrow \mathbb{K}_1 (1 - \delta),
\quad \mathbb{K}_2 \rightarrow \mathbb{K}_2 (1 - \delta) ,
\end{eqnarray}
where $\delta$ (with $\delta >0$) is a phenomenological parameter that breaks
an appropriate lattice symmetry for the PCO. 
This modification of the ring hoppings captures the spatial modulation
of the energy in the system when the PCO is present. 
This phenomenological way of introducing the PCO is very similar in spirit to Henley's 
approach \cite{Henley87} to the order by disorder, where a phenomenological interaction is introduced into the energy or the free energy to model the ground state 
selection due to the quantum fluctuation.

\begin{figure}[t]
\subfigure[\, $\delta = 0$]{\includegraphics[width=4.2cm]{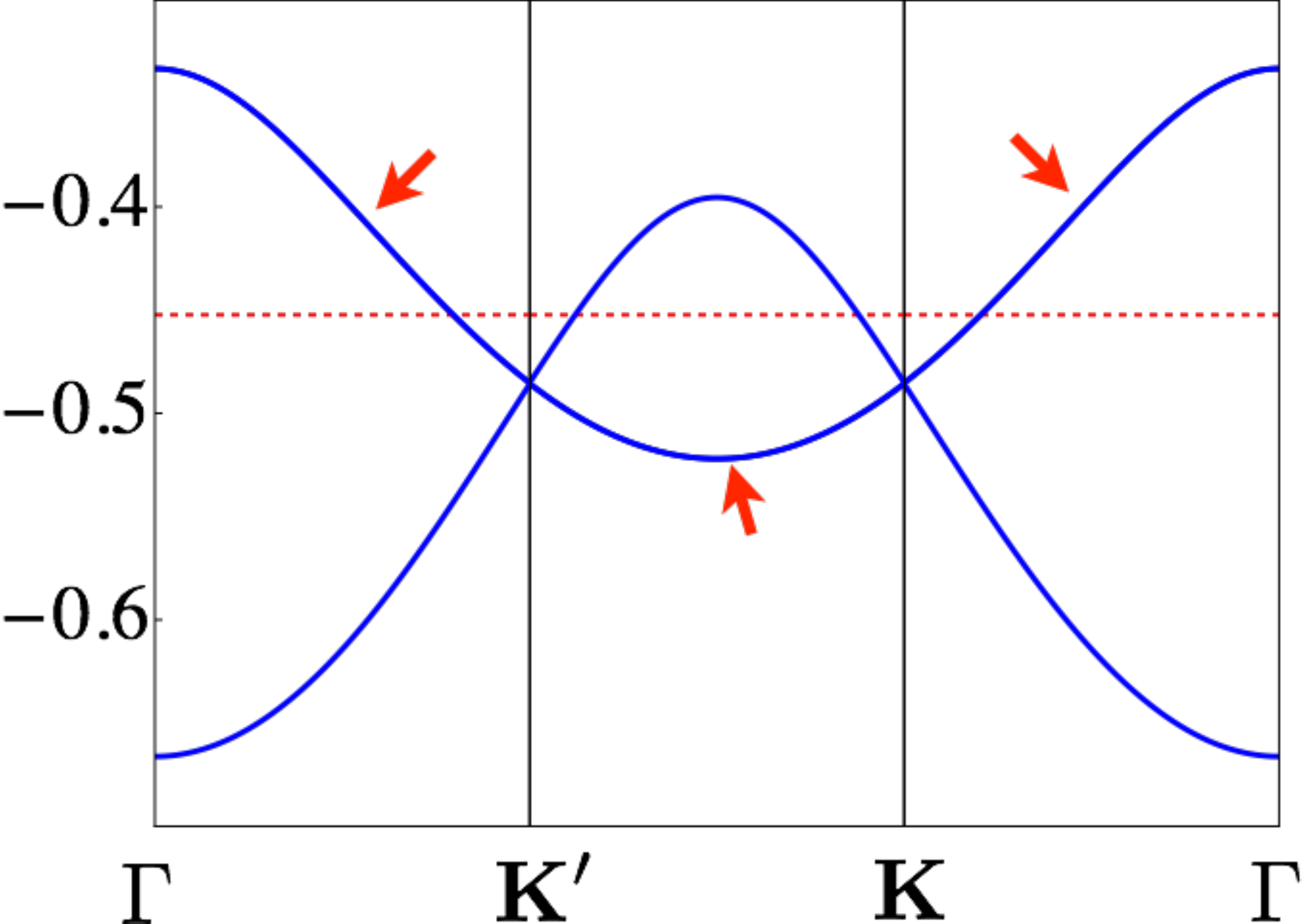}}
\subfigure[\, $\delta =0.3$ ]{\includegraphics[width=4.2cm]{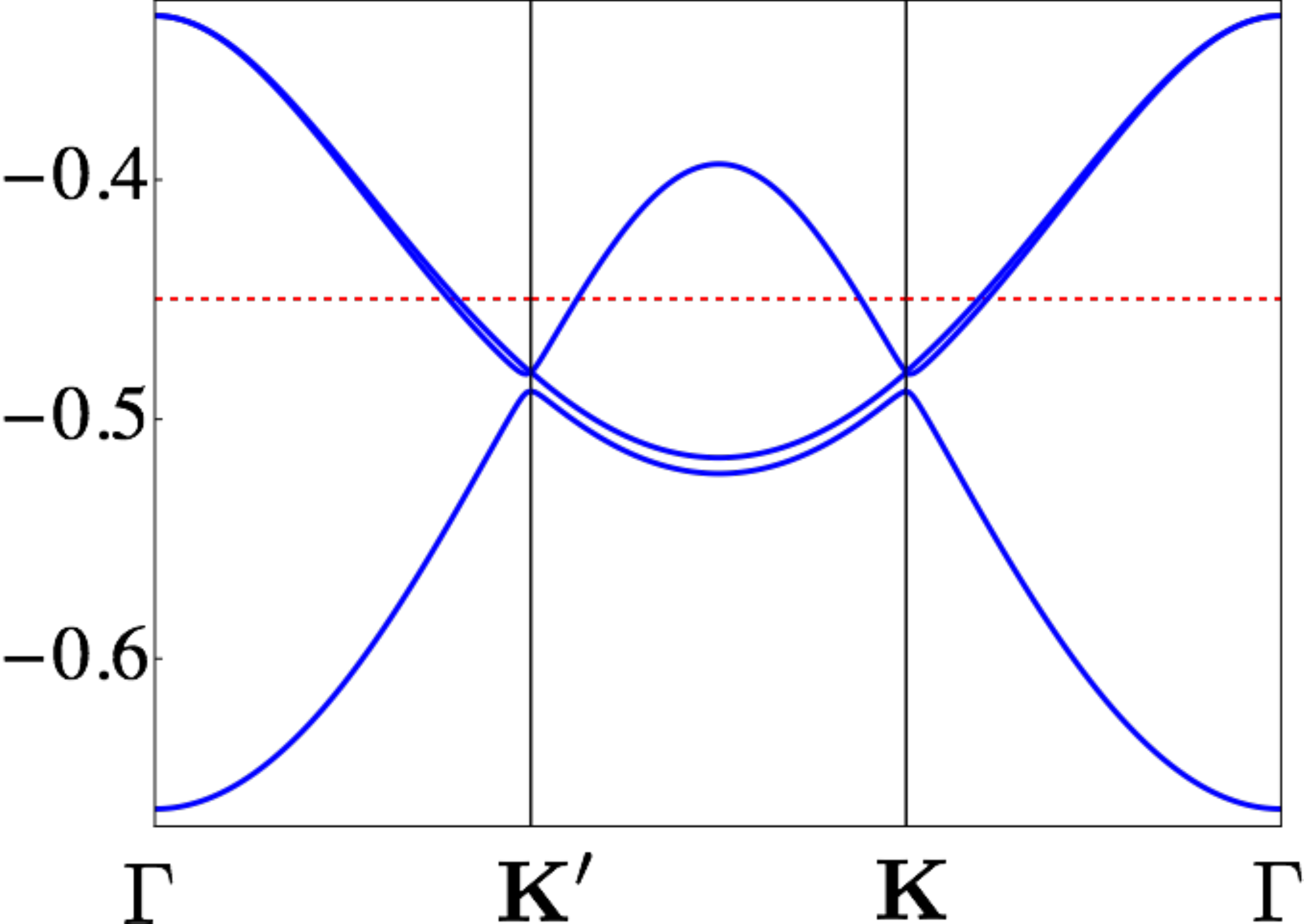}}
\subfigure[\, $\delta =0.7$]{\includegraphics[width=4.2cm]{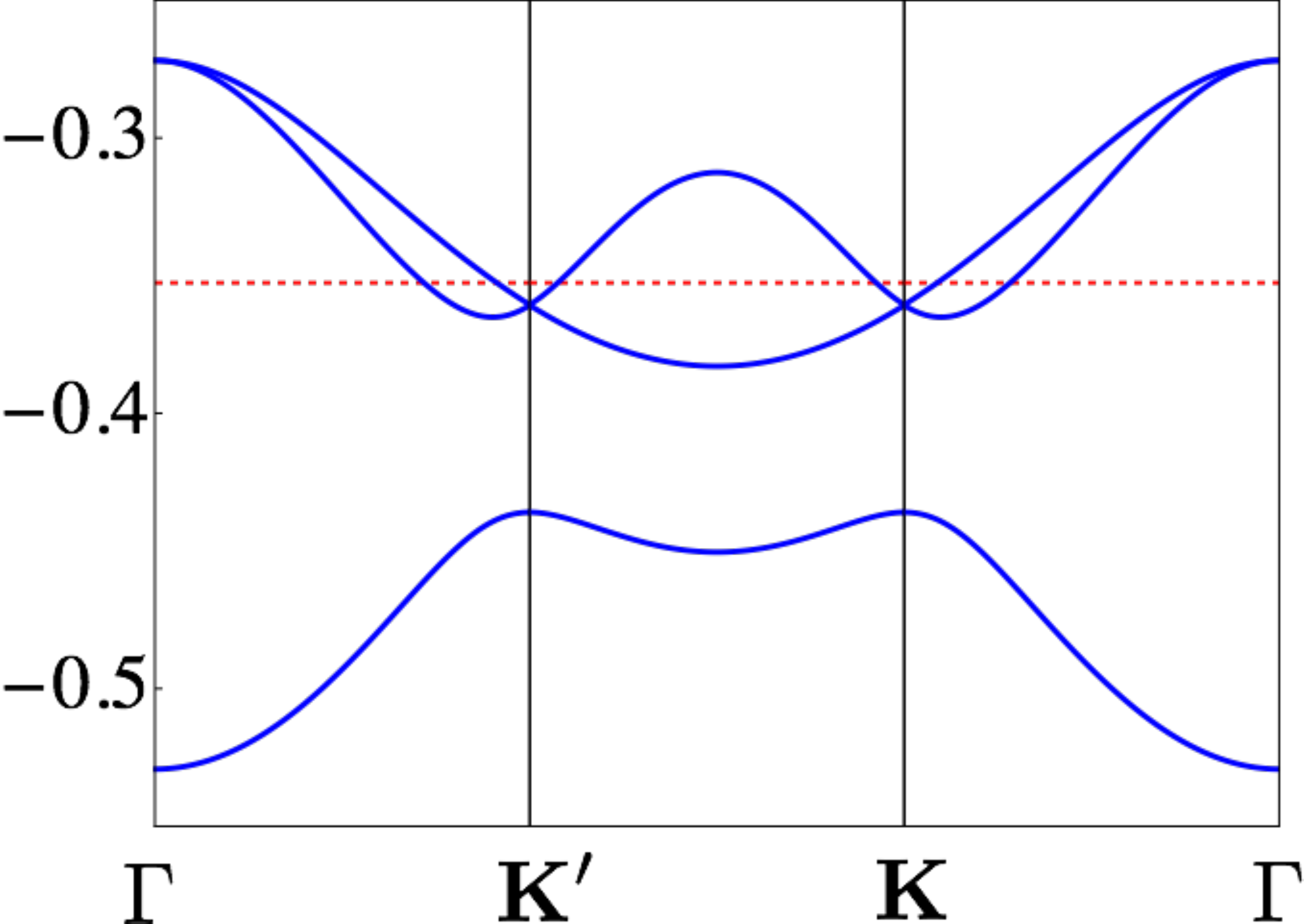}}
\hspace{0.7cm}
\subfigure[\, Brioullin zones]{\includegraphics[width=3.4cm]{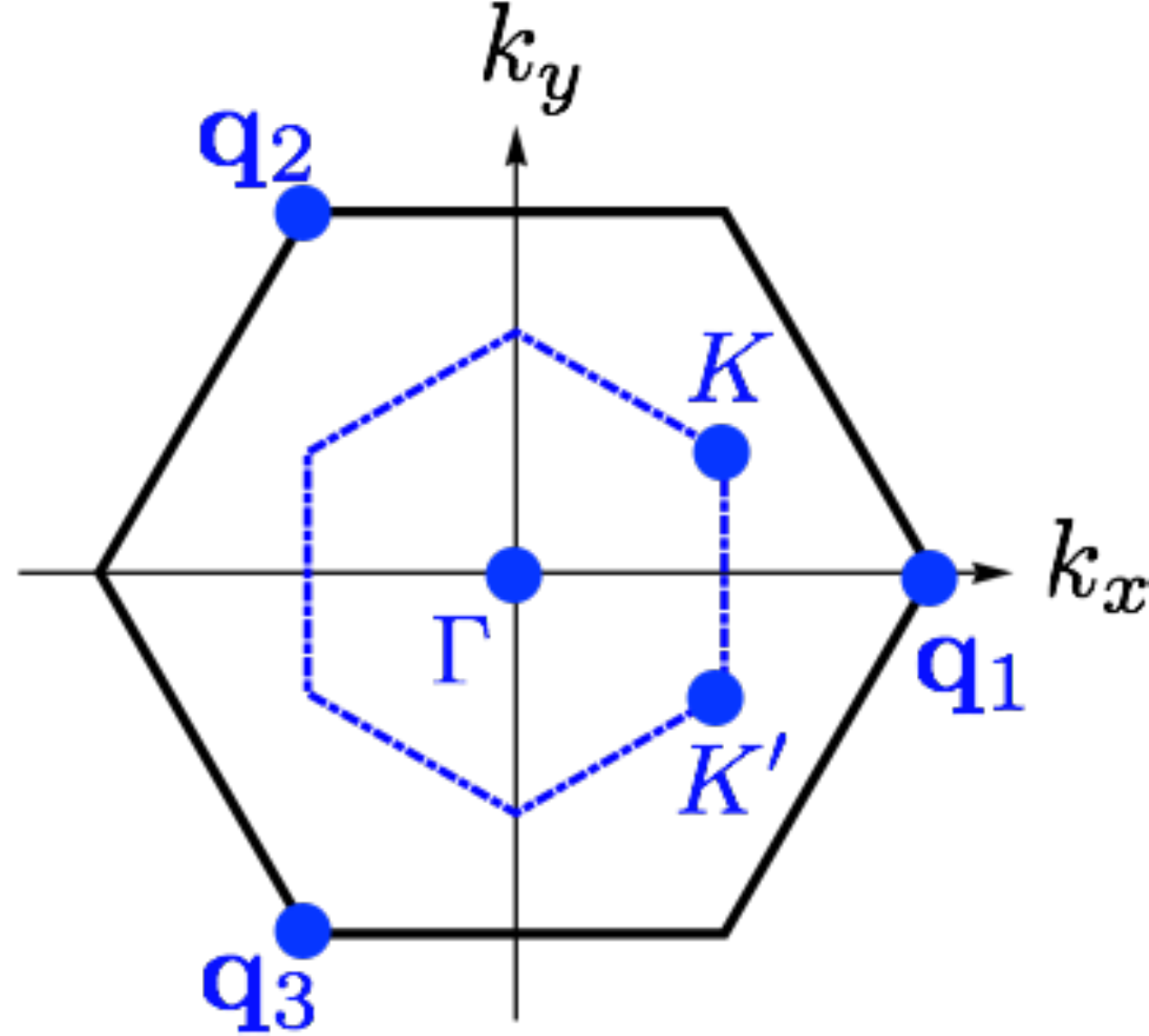}}
\caption{(Color online.) (a-c) The evolution of the spinon sub-bands as $\delta$ is varied.  
In the figure, we choose $\mathbb K_1 =4, \mathbb K_2 =1$ and $t_1=1, t_2 =0.5$. 
The (red) dashed line is the Fermi energy of the spinons. The energy unit is set to $t_1$.
The (red) arrow indicates a 2-fold degeneracy. 
The details of the band structures are discussed in the text. 
(d) The large and small hexagons define the Brioullin
zones of the Kagome lattice (BZ1) and the ETL (BZ2),
respectively. 
Setting the Kagome lattice constant to unity, we have 
${\bf K} = (\frac{2\pi}{3}, \frac{2\pi}{3\sqrt{3} }), 
{\bf K}' =(\frac{2\pi}{3}, -\frac{2\pi}{3\sqrt{3} })$
and ${\bf q}_1 = (\frac{4\pi}{3},0), {\bf q}_2 = ( - \frac{2\pi}{3}, \frac{2\pi}{\sqrt{3}}),
{\bf q}_3 =   ( - \frac{2\pi}{3},-\frac{2\pi}{\sqrt{3}})$.
}
\label{fig12}
\end{figure} 

We now solve the combined Hamiltonian of 
$\bar{H}_{\text{s}}$ and $\bar{H}_{\text{ring}}$
with the modified ring hoppings self-consistently. 
This self-consistent approach is expected to
underestimate the PCO for any fixed $\delta$ and thus 
underestimates the reconstruction of the spinon 
band structure due to the PCO. 
Nevertheless, to understand the generic features of the spinon band structure 
reconstruction in the presence of the PCO, we can simply vary the phenomenological parameter $\delta$ and study the spinon band structure from this self-consistent approach. 

The evolution of the mean-field spinon band structure is depicted in Fig.~\ref{fig12}.
When $\delta = 0$, there is no PCO and the symmetry 
of the Kagome lattice is preserved. 
The spinon band structure contains 3 bands in the 1st Brioullin zone of the Kagome lattice 
(BZ1 in Fig.~\ref{fig12}d). The 3 spinon bands are well separated in energy and 
have no direct nor indirect overlap, and we can simply focus on the lowest band
as the spinons only fill half of the lowest band. 
So in Fig.~\ref{fig12}, we only need to plot the evolution of the lowest spinon band.
In Fig.~\ref{fig12}a, we have further folded the lowest spinon band 
 onto the 1st Brioullin zone of the ETL (BZ2 in Fig.~\ref{fig12}d)
 and obtain 3 spinon sub-bands.  
We use ``{\it spinon sub-bands}'' to refer the spinon bands plotted in the BZ2 of the ETL. 
 
In the presence of the PCO for a finite $\delta$, the system has 9 sublattices (see Fig.~\ref{fig5}), the 3 spinon bands at $\delta=0$ are further split into 9 spinon {\it sub-bands}, 
and the lowest spinon band at $\delta=0$ is split into 3 spinon {\it sub-bands}. 
 
We now explain the evolution of the spinon bands as the PCO is enhanced by increasing
the variational parameter $\delta$.  
At $\delta=0$, there is no PCO, and the 2nd and the 3rd spinon sub-bands 
touch at the zone boundary of the BZ2. 
A finite $\delta$ creates the PCO and breaks the translational symmetry of the 
Kagome lattice. We have $\tilde{t}_1 > \tilde{t}_1'$
and $\tilde{t}_2 > \tilde{t}_2'$ as previously expected.  
The band touching of the 2nd and the 3rd spinon sub-bands 
 at the zone boundary of the BZ2
 is lifted by the level repulsion (see Fig.~\ref{fig12}b).   
A {\it direct band gap} is created between the lowest spinon sub-band and upper
spinon sub-bands. 
We emphasize this feature is generic and is not specific to the 
ring hoppings and electron hoppings that are chosen in Fig.~\ref{fig12}
and we have also explicitly checked many other parameter choices. 
As the parameter $\delta$ is further increased and the PCO becomes even stronger,
the direct band gap gets larger and larger, 
and eventually the lowest spinon sub-band is fully separated from the other sub-bands 
by a full band gap (see Fig.~\ref{fig12}c). Therefore, the band gap of the spinons 
is set by the stiffness of the PCO. 

Even though the PCO enlarges the unit cell from 3 sites of the Kagome lattice to 
9 sites of the ETL, the spinon Fermi surface always exists.
This is because the number of electrons or spinons per unit cell is 3 with the PCO. 
Because of the direct band gap, the lowest spinon sub-band is completely filled by the spinons which comprise 2/3 of the total spinon number, and 
the remaining 1/3 of spinons partially fill the upper 2 sub-bands and 
give rise to the spinon Fermi surfaces. 
Therefore, the internal U(1) gauge field is expected to be in the deconfining phase \cite{SungSik2008,Hermele2004}, and we obtain U(1) QSL for the ground state.

\section{Thermal transition and spin susceptibility}
\label{sec4}

Because the PCO breaks the lattice symmetry, this implies that there exists a
thermal phase transition at a finite temperature which destroys the
PCO and
restores the lattice symmetry. This thermal transition is expected to occur at 
$T^{\ast} \sim \mathcal{O}( \mathbb{K}_1) = \mathcal{O} (t_1^3/V_2^2) $
(because $\mathbb{K}_1 \gg \mathbb{K}_2 $) when the 
local electron resonance in the elementary hexagons loses the 
quantum phase coherence. 

Based on the understanding of the spinon band structure in Sec.~\ref{sec3}, 
we describe the behavior of 
the spin susceptibility in the low temperature regime with the PCO ($T< T^{\ast}$) and in the high temperature regime without the PCO ($T>T^{\ast}$). 
Since the U(1) QSL ground state has spinon Fermi surfaces, 
we expect a finite (Pauli-like) spin susceptibility in the zero temperature limit. 
At finite temperatures, one should recover the Curie-Weiss law for the 
spin susceptibility. What are the Curie constant and the Curie-Weiss temperature 
that characterize the Curie-Weiss law for $T<T^{\ast}$ and $T> T^{\ast}$? 
The Curie constant measures the number of the {\it active} local moments. 
Let us now consider the local moment formation regime for $T<T^{\ast}$. 
As long as the PCO is not destroyed by thermal fluctuations which is the case for $T<T^{\ast}$,
the direct band gap between the lowest spinon sub-band and upper sub-bands would 
be present, and the lowest spinon sub-band is fully filled by 2/3 of the spinon numbers.
A fully-filled spinon is {\it inert} to an external magnetic field,
and thus, only the 1/3 of the spinons from the partially 
filled upper sub-bands contribute to the 
local moment, which comprise 1/3 of the total number of electrons in the system. 
Therefore, the low temperature Curie constant in a DC susceptibility measurement
for $T<T^{\ast}$ is 
\begin{equation}
{\mathcal C}^{\text L} = \frac{g^2 \mu_B^2 s(s+1)}{3k_{\text B}} \frac{N_{\Delta}}{3} 
\end{equation}
where $g\approx 2$ is the Land\'{e} factor \cite{Sheckelton12,Sheckelton14}, 
$s={1}/{2}$, and 
$N_\Delta$ is the total number of up-triangles in the system.   
From the electron filling fraction, we know $N_{\Delta}$ equals the total electron number $N_e$.  Because only 1/3 of the total spins are responsible for the low-temperature magnetic properties, the Curie constant is 
only 1/3 of the one at very high temperatures where all the 
electron spins are supposed to be active. 

For $T>T^{\ast}$, the PCO is destroyed by thermal fluctuation, and the direct band gap
between the lowest spinon sub-band and the upper spinon sub-bands is closed.  
All the localized electrons are active and contribute to the local moment, and thus the Curie constant 
in this high temperature regime is 
\begin{equation}
{\mathcal C}^{\text H} = \frac{g^2 \mu_B^2 s(s+1)}{3k_{\text B}}  {N_{\Delta}},
\end{equation}
which is 3 times the low temperature one, ${\mathcal C}^{\text L}$. 

As for the Curie-Weiss temperature, it is hard to make a quantitative 
prediction from the spinon Fermi surface. 
But it is noted that the Curie-Weiss temperature is roughly set by the 
bandwidth of the {\it active} spinon bands in the QSL phase.
At low temperature PCO phase, the active spinon bands are the partially-filled 
upper spinon sub-bands on the ETL (see Fig.~\ref{fig12}c). 
As one can see from Fig.~\ref{fig12}c, 
the bandwidth of the active spinon bands is significantly reduced when the PCO is present
compared to Fig.~\ref{fig12}a when the PCO is absent. 
As a result, we expect a much reduced Curie-Weiss temperature
in the presence of the PCO at $T< T^{\ast}$
compared to the case in the absence of the PCO at $T>T^{\ast}$. 
In the absence of the PCO at $T>T^{\ast}$, 
as all the spinon sub-bands are active, the Curie-Weiss temperature 
is set by the total spinon bandwidth in Fig.~\ref{fig12}a.
Finally, we want to point out that the double Curie regimes
in the spin susceptiblity is a finite temperature property and 
independent from whether the spin ground state is a spinon Fermi surface
U(1) QSL or not. It is the PCO that reconstructs the spin states
and leads to the double Curie regimes. In the Appendix~\ref{asec2},
we provide a complementary explanation of the double Curie regimes 
from the spin state reconstruction in the strong PCO regime.

\section{Discussion}
\label{sec5}

\subsection{Applications to LiZn$_2$Mo$_3$O$_8$}
\label{sec5A}

As we discuss in Sec.~\ref{sec4}, the CMI with the PCO 
provides two Curie-Weiss regimes in spin susceptibility. 
Armed with these results, 
we here propose that the Mo system in LiZn$_2$Mo$_3$O$_8$ 
may be in the CMI with the PCO at low temperatures. 
Because the PCO triples the unit cell, the thermal transition 
at $T^{\ast}$ is found to be first order in a Landau theory analysis 
for a clean system \cite{GangChen2014b}. 
In reality, LiZn$_2$Mo$_3$O$_8$ is influenced by various disorders or 
impurities (e.g. the mixed Li/Zn sites 
and mobile Li ions) \cite{Sheckelton12}. For example, impurities 
would broaden the charge ordering transition \cite{McMillan75}. 
This may explain why a sharp transition is not observed 
in the experiments \cite{Sheckelton12}. 
Nevertheless, the experiments do observe a peak around 100K in 
heat capacity \cite{Sheckelton12} which might be related to the smeared-out 
phase transition. 

Based on the fact that there is no obvious ordering down to $\sim$ 0.1K for LiZn$_2$Mo$_3$O$_8$ and the apparent gapless spin excitation in neutron scattering \cite{Sheckelton12,Sheckelton14,Mourigal14}, we further propose that the system is in
the U(1) QSL with spinon Fermi surfaces (as well as the PCO)
of the CMI that is obtained in Sec.~\ref{sec3B}.  
With the spinon Fermi surfaces, we expect the usual behaviors of a 2D U(1) QSL with 
spinon Fermi surfaces would show up. That is, the specific heat $C_v \sim T^{2/3}$,
and a Pauli-like spin susceptibility in the low temperature limit \cite{Lee05}. 
The crossover in the behaviour of the spin susceptibility 
from the local moment Curie-Weiss regime to the 
Pauli-like behaviour is expected to happen at the 
temperature set by the bandwidth of active spinon bands (see Sec.~\ref{sec3B}), 
or equivalently, by the low-temperature Curie-Weiss temperature below $T^{\ast}$. 
This crossover temperature should be very low because of the suppressed Curie-Weiss temperature at low temperatures. As a result, the Pauli-like spin susceptibility may be smeared out by various extrinsic factors like local magnetic impurities at very low temperatures. Likewise, even though the $C_v/T$ experiences a upturn 
below 10K in the absence of external magnetic fields,  
it is likely that the nuclear Schottky anomalies may complicate the specific heat data. 

On the other hand, the apparently gapless spectrum of the spin excitations in the inelastic 
neutron scattering measurement \cite{Mourigal14} is certainly 
consistent with the gapless spinon Fermi surface of our U(1) QSL. Moreover,
the measurements of relaxation rate from both NMR ($1/(T_1 T)$) and 
 $\mu$SR ($\lambda T^{-1}$) also indicate gapless spin-spin correlations \cite{Sheckelton14}. In our U(1) QSL, the reduction of the spinon bandwidth due to the PCO increases the density of the low-energy magnetic excitations. This would lead to a 
low-temperature upturn of the spin-lattice relaxation, which is in fact observed in 
NMR and $\mu$SR experiments \cite{Sheckelton14}. 

A direct measurement of the PCO at low temperatures is crucial for our theory. 
To this end, a high resolution X-ray scattering measurement and 
NQR (nuclear quadrupole resonance) can be helpful. 
Moreover, the presence of local quantum entanglement within the resonant
hexagon may be probed optically by measuring the local exciton excitations. 
Furthermore, if the system is in a U(1) QSL with a spinon Fermi surface, 
the low-temperature thermal conductivity can be an indirect probe of
the low-energy spinon excitation, and a direct measurement of the 
correlation of the emergent U(1) gauge field might be possible because 
the strong spin-orbit coupling of the Mo atoms can enhance the coupling
between the spin moment and the spin texture \cite{Lee13}.

A previous work on LiZn$_2$Mo$_3$O$_8$ has proposed a theory based on 
varying spin exchange interaction from the emergent lattice that is caused by the lattice distortion \cite{Flint13}. In contrast, our work here is based on the electron degrees of freedom and their interactions. In Appendix.~\ref{asec1}, we clarify the 
 underlying quantum chemistry of LiZn$_2$Mo$_3$O$_8$ and
 justify the extended Hubbard model of Eq.~\eqref{eq1}.

\subsection{Other Mo based cluster magnets}
\label{sec5B}

\begin{table}[t]
\centering
\begin{tabular}{lccccc}
\hline\hline 
& [Mo-Mo]$_{\text u}$ & [Mo-Mo]$_{\text d}$ & $\lambda$ & e$^-$/Mo$_3$ & Ref
\\
\hline\hline
LiZn$_2$Mo$_3$O$_8$ & 2.6\AA & 3.2\AA & 1.23 & 7 & [\onlinecite{Sheckelton12}]
\\
Li$_2$InMo$_3$O$_8$ & 2.54\AA & 3.25\AA & 1.28 & 7 & [\onlinecite{Gall13}]
\\
ScZnMo$_3$O$_8$ & 2.58\AA & 3.28\AA  & 1.27 & 7 & [\onlinecite{TORARDI85}]
\\
\hline\hline
\end{tabular}
\caption{Mo-Mo bond lengths, 
 anisotropic parameters ($\lambda$) 
and number of electrons per Mo$_3$O$_{13}$ cluster 
 for three different cluster magnets. 
The electron number is counted from stoichiometry.}
\label{tab3}
\end{table}

The compounds that incorporate the Mo$_3$O$_{13}$ cluster unit
represent a new class of magnetic materials called ``cluster magnets''.  
Several families of materials, such as M$_2$Mo$_3$O$_8$ 
(M = Mg, Mn, Fe, Co, Ni, Zn, Cd), LiRMo$_3$O$_8$ (R = Sc, Y, In, Sm, Gd, Tb, Dy, Ho, Er, Yb) and other related variants\cite{McCarroll57,McCarroll77,Gall13,TORARDI85}, 
fall into this class. The magnetic properties of most materials 
have not been carefully studied so far. 
In Tab.~\ref{tab3}, we list three cluster magnets with odd 
number of electrons in the Mo$_3$O$_{13}$ cluster unit. 
We introduce a phenomenological parameter $\lambda$ to characterize
the anisotropy of the Mo Kagome lattice, which is defined as 
the ratio between inter-cluster (or down-triangle) and intra-cluster (or up-triangle) 
Mo-Mo bond lengths,
\begin{equation}
\lambda = \frac{\text{[Mo-Mo]}_{\text d}}{\text{[Mo-Mo]}_{\text u}}. 
\end{equation}
Large anisotropy tends to reduce the interaction $V_2$ and increase the
hopping $t_1$ so that the systems are more likely to be in the regime where
the electron is only localized on the up-triangle while the electron number on the 
down triangle is strongly fluctuating (see Fig.~\ref{fig1}a). 
In such a regime, there is no PCO, each localized electron on the up-triangle forms a local spin-1/2 moment, and these local spin-1/2 moments form a triangular lattice. If the system is in the weak Mott regime like the organics \cite{itou08,Shimizu03}, the spin ground state is expected to be the U(1) QSL with a spinon Fermi surface \cite{GangChen2014b}. 

As shown in Tab.~\ref{tab3}, Li$_2$InMo$_3$O$_8$ has a larger anisotropic
parameter than LiZn$_2$Mo$_3$O$_8$. 
Unlike LiZn$_2$Mo$_3$O$_8$ that has two Curie-Weiss regimes, the spin susceptibility of 
Li$_2$InMo$_3$O$_8$ is instead characterized by one Curie-Weiss temperature $\Theta_{\text{CW}} = -207$K
down to 25K.\cite{Gall13}
Moreover, the Curie constant is consistent with one local spin-1/2 moment 
in each up-triangle. 
Below 25K, the spin susceptibility of Li$_2$InMo$_3$O$_8$ 
saturates to a constant, which is consistent with the Pauli-like spin susceptibility 
for a spinon Fermi surface U(1) QSL. Besides the structural and spin susceptibility data,
very little is known about Li$_2$InMo$_3$O$_8$.  Thus,
more experiments are needed to confirm the absence of magnetic ordering in 
Li$_2$InMo$_3$O$_8$ and also to explore the magnetic properties of ScZnMo$_3$O$_8$
and other cluster magnets.

\section{Acknowledgements}

GC thanks A. Essin, A. Burkov, L. Balents, M. Hermele, Fuchun Zhang for helpful discussion,
and especially T. McQueen and P.A. Lee for email correspondence and conversation.
HYK and YBK are supported by the NSERC, CIFAR, and Centre for Quantum Materials 
at the University of Toronto. GC is supported by the starting-up funding of 
Fudan University (Shanghai, People's Republic of China) and Thousand-Youth 
Talent program of People's Republic of China. GC acknowledges NSF grant no.~PHY11-25915 
for supporting the visitor program at the Kavli Institute for Theoretical Physics during 
the the workshop ``Frustrated Magnetism and Quantum Spin Liquids'' in October 2012, 
where the current work was inspired and initiated.  

\appendix

\section{Molecular orbitals and the Hubbard model}
\label{asec1}

\begin{figure}[t]
{\includegraphics[width=8cm]{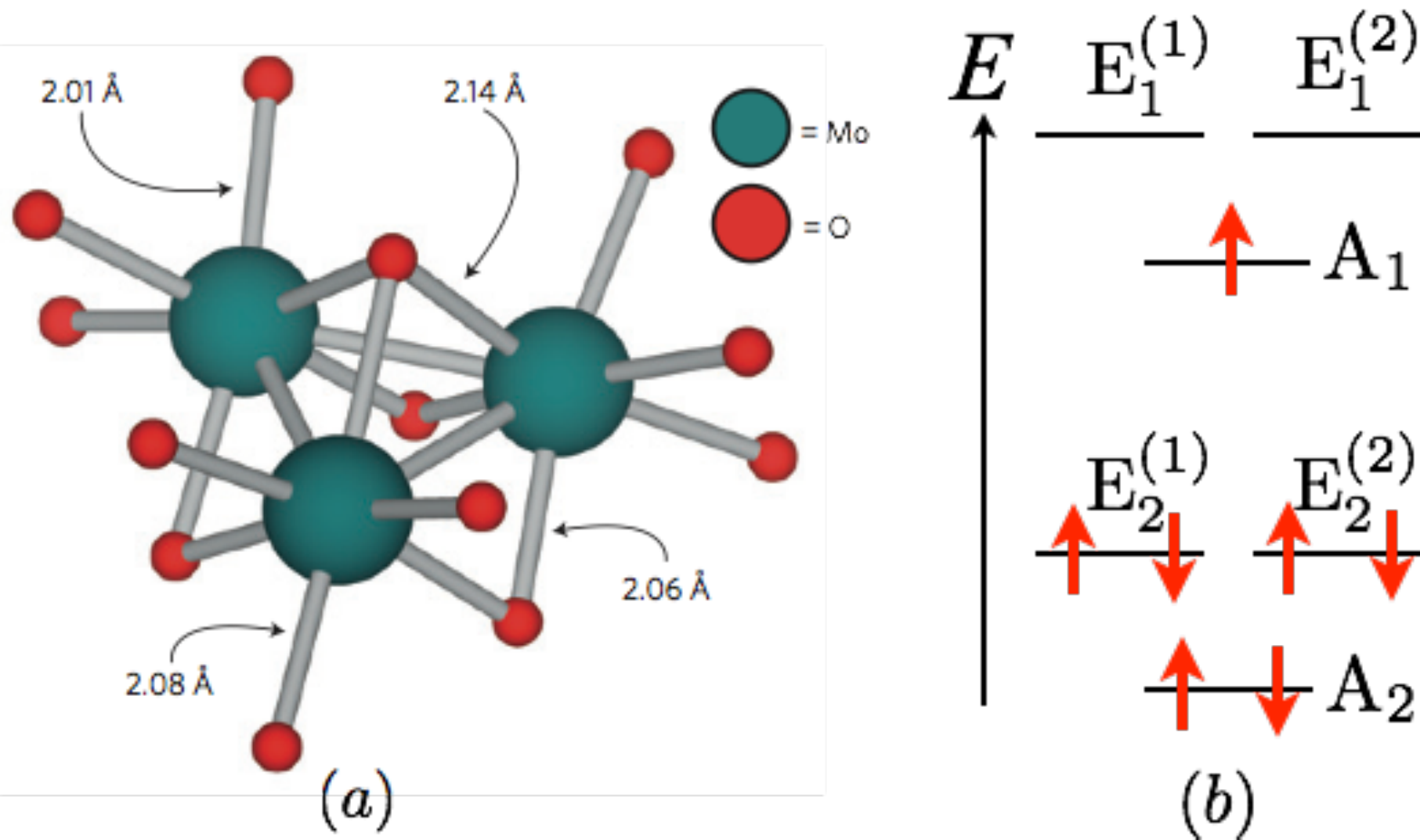}}
\caption{(Color online.) 
(a) The Mo$_3$O$_{13}$ cluster (adapted from Ref.~\onlinecite{Sheckelton12}).
(b) The schematic energy level diagram of the molecular orbitals for a single Mo$_3$O$_{13}$ cluster. 
The molecular orbitals are classified according to the irreducible representations of the 
C$_{3v}$ point group of the cluster\cite{Note1}.
The unfilled molecular orbitals at high energies are not shown. 
}
\label{afig1}
\end{figure}

As suggested by Refs.~\onlinecite{Cotton64,Sheckelton12}, the Mo electrons in an 
isolated Mo$_3$O$_{13}$ cluster form molecular orbitals
because of the strong Mo-Mo bonding. Among the 7 valence electrons in 
the cluster, 6 of them fill the lowest three molecular orbitals \{A$_2$, 
E$_2^{(1)}$, E$_2^{(2)}$\} in pairs, and the seventh electron remains 
unpaired in a totally symmetric A$_1$
molecular orbital with equal contributions from all three Mo atoms (see Fig.~\ref{afig1}). 

We first consider the molecular orbital states in the group
\{A$_1$, E$_1^{(1)}$, E$_1^{(2)}$\}. 
This group can be described by a linear combination of 
an atomic state $|\psi_1\rangle$ at each Mo site (which is in turn
a linear combination of five $4d$ atomic orbitals).
\begin{eqnarray}
|{\text A}_1 \rangle &=& \frac{1}{\sqrt{3}} 
\big[ | {\psi}_1 \rangle_{\text A} +  | {\psi}_1 \rangle_{\text B}
+  | {\psi}_1 \rangle_{\text C}\big],
\label{eqA2}\\
|{\text E}_1^{(1)} \rangle &=& \frac{1}{\sqrt{3}} \big[ | {\psi}_1 \rangle_{\text A} + 
e^{ i \frac{2\pi}{3}} | {\psi}_1 \rangle_{\text B}
+ e^{ - i \frac{2\pi}{3}} | {\psi}_1 \rangle_{\text C}\big],
\label{eqE2a}
\\
|{\text E}_1^{(2)}  \rangle &=& \frac{1}{\sqrt{3}} \big[ | {\psi}_1 \rangle_{\text A} +
e^{- i \frac{2\pi}{3}}   | {\psi}_1 \rangle_{\text B}
+ e^{ i \frac{2\pi}{3}} | {\psi}_1 \rangle_{\text C}\big],
\end{eqnarray}
where $\mu $($= \text{A,B,C}$) labels the three Mo sites in the cluster 
and the atomic state $| {\psi}_1 \rangle_{\mu}$ is the contribution
from the Mo atom at $\mu$. The atomic states $| {\psi}_1 \rangle_{\mu}$ 
at different Mo sites are related by the 3-fold rotation about the center of the cluster. 
Likewise, the fully-filled \{A$_2$, E$_2^{(1)}$, E$_2^{(2)}$\} and other unfilled 
molecular orbitals at higher energies are constructed from 
the atomic state $|\psi_2\rangle$ and other atomic states
$|\psi_j\rangle$ ($j=3,4,5$), respectively. Here,
the {\it atomic states} $\{ | {\psi}_j \rangle_{\mu} \}$ ($j=1,2,3,4,5$)
represent a {\it distinct} orthonormal basis from the five $4d$ {\it atomic orbitals} 
that are the eigenstates of the local Hamiltonian of the MoO$_6$ octahedron.  

We group the molecular orbitals based on the atomic state from which they are 
constructed. In this classification, for example, \{A$_1$, E$_1^{(1)}$, E$_1^{(2)}$\}
fall into one group while \{A$_2$, E$_2^{(1)}$, E$_2^{(2)}$\} fall into another group
as they are constructed from two different atomic states. 

In LiZn$_2$Mo$_3$O$_8$, the different molecular orbitals 
of the neighboring clusters Mo$_3$O$_{13}$ overlap and form molecular bands. 
To understand how the molecular orbitals overlap with each other, we 
consider the wavefunction overlap of different atomic states $|\psi_j\rangle$. 
Since the down-triangle has the same point group symmetry as the up-triangle in LiZn$_2$Mo$_3$O$_8$,
the wavefunction overlap of the atomic states in
the down-triangles should approximately resemble the one in the up-triangles.   
More precisely, the wavefunction of the atomic state (e.g. $|\psi_1\rangle$)
has similar lobe orientations both inward into and outward from the Mo$_3$O$_{13}$
cluster, with different spatial extensions due to the asymmetry 
between up-triangles and down-triangles.  
Consequently, the orbital overlap between the molecular orbitals from the same group
 is much larger than the one 
between the molecular orbitals  
from the different groups.  
Therefore, each molecular band cannot be formed by one single molecular orbital
but is always a strong mixture of the three molecular orbitals in the same group. 

We now single out the three molecular bands that are primarily formed by 
the group of \{A$_1$, E$_1^{(1)}$, E$_1^{(2)}$\} molecular orbitals. 
There are four energy scales associated with these three
molecular orbitals and bands:  \\
(1) the energy separation $\Delta E $ 
between the  \{A$_1$, E$_1^{(1)}$, E$_1^{(2)}$\}  group
 and other groups of orbitals (both filled and unfilled), \\
(2) the {\it total} bandwidth $W$ of the three molecular bands
formed by the \{A$_1$, E$_1^{(1)}$, E$_1^{(2)}$\} molecular orbitals, \\
(3) the intra-group interaction between two electrons on 
any one or two orbitals of the \{A$_1$, E$_1^{(1)}$, E$_1^{(2)}$\} group, \\
(4) the inter-group interaction between the electron 
on an orbital of the \{A$_1$, E$_1^{(1)}$, E$_1^{(2)}$\} group
and the other electron on an orbital of a different group.
It is expected, from the previous wavefunction overlap argument, that the inter-group 
interaction is much weaker than the intra-group interaction 
and thus can be neglected at the first level of approximation. 

In this paper, we assume that the energy separation $\Delta E$
is larger than the total bandwidth $W$ and the intra-group interaction. 
In this regime, 
the large $\Delta E$ separates these three molecular bands 
from other molecular bands (both filled and unfilled) so that 
the fully filled \{A$_2$, E$_2^{(1)}$, E$_2^{(2)}$\} orbitals remain
fully-filled and the unfilled molecular orbitals remain unfilled even after they
form bands. 
Moreover, the large $\Delta E$ also prevents a band-filling reconstruction 
due to the interaction (in principle, the system can gain interaction energy by 
distributing the electrons evenly among different groups of orbitals). 
Therefore, we can ignore both the fully-filled and unfilled molecular bands
and just focus on the three partially filled bands. 
It also means one will have to consider three-band model 
with all of \{A$_1$, E$_1^{(1)}$, E$_1^{(2)}$\} orbitals
on the triangular lattice formed by the Mo$_3$O$_{13}$ clusters. 
In this case, alternatively one could simply consider  
atomic states as the starting point. Then the relevant model
would be a single-band Hubbard model based on the atomic state $|\psi_1\rangle$ 
at each Mo site of the anisotropic Kagome lattice.
We take the latter approach in this paper.
Finally, since only one atomic state $|\psi_1\rangle$ is involved at 
each Mo site, the orbital angular momentum of the electrons are trivially quenched
so that we can neglect the atomic spin-orbit coupling
at the leading order\cite{William14}. 

The corresponding single-band Hubbard model is given by Eq.~\eqref{eq1},
where we include the on-site and nearest-neighbor electron interactions. 
Now it is clear that the physical meaning of the electron operator $c^\dagger_{i\sigma}$ 
($c^{\phantom\dagger}_{i\sigma}$) in Eq.~\eqref{eq1} 
is to create (annihilate) an electron on the 
state $|\psi_1 \rangle_i$ with spin $\sigma$ at the Kagome lattice site $i$.

\section{Local moments in the strong PCO limit}  
\label{asec2}

In Sec.~\ref{sec4}, we show the PCO 
reconstructs the spinon band structure and 
provide a possible explanation of the 
doule Curie regimes and 1/3 spin susceptibility in LiZn$_2$Mo$_3$O$_8$. 
As we point out that the 
doule Curie regimes and 1/3 spin susceptibility
are finite temperature properties and independent from whether the 
spin ground state has a spinon band or not, in this section, we 
onsider an alternative and complementary strong coupling regime
in which the PCO is strong such that the 3 resonating electrons are 
almost fully localized in the resonating hexagon and form the local moments 
which then interact with each other. 
As far as the local moment physics is concerned, the 
regime considered here is   
equivalent to the intermediate PCO regime in 
Sec.~\ref{sec4}~\cite{Lee05,Motrunich05,Motrunich06}. 

To elucidate the nature of the local moments in each resonating hexagon,  
it is sufficient to isolate a single resonating hexagon and understand
the local quantum entanglement among the 3 resonating electrons. 
In this subsection, we consider two local interactions on the hexagon. 
The first interaction is already given in Eq.~\eqref{eq65} which is
the electron ring hopping model. The second interaction is the 
AFM exchange interaction between the electron 
spins. Since the electrons are always separated from each other by one lattice site, 
the AFM exchange is between the next nearest neighbors in the hexagon,
\begin{equation}
H_{\text{ex}}^0 =  J\sum_{\langle\langle ij\rangle\rangle}  
n_i n_j  ({\bf S}_i \cdot {\bf S}_j - \frac{1}{4} ),
\label{eq74}
\end{equation} 
where $i,j$ are the lattice sites that refer to the 6 vertices of the 
resonating hexagon (see Fig.~\ref{fig1}), $n_i$ is the electron occupation 
number at the site $i$ and ${\bf S}_i$ is the spin-1/2 operator of 
the electron spin at the site $i$. Because the electron position is not 
fixed in the hexagon, the AFM interaction is active only when both relevant 
sites are occupied by the electrons and we need to include $n_i$ into 
the exchange interaction. The full Hamiltonian for an individual 
resonating hexagon plaquette is composed of the above two interactions,
\begin{eqnarray}
H_{\text p} &=& - \sum_{\alpha\beta\gamma} 
                    \big[\mathbb{K}_1 ( c^{\dagger}_{1\alpha} c^{\phantom\dagger}_{6\alpha} 
                                 c^{\dagger}_{5\beta} c^{\phantom\dagger}_{4\beta}  
 c^{\dagger}_{3\gamma} c^{\phantom\dagger}_{2\gamma} + h.c.)
\nonumber \\
 &&       
+ \mathbb{K}_2 ( c^{\dagger}_{1\alpha} c^{\phantom\dagger}_{2\alpha} 
                                 c^{\dagger}_{3\beta} c^{\phantom\dagger}_{4\beta}  
                                  c^{\dagger}_{5\gamma} c^{\phantom\dagger}_{6\gamma}  
+ h.c. ) \big] + H^0_{\text{ex}}.
\end{eqnarray}
Based on the perturbative values $\mathbb{K}_1 = 6t_1^3/V_2^2,
\mathbb{K}_2= 6t_2^3/V_1^2$ and the fact that $t_1 > t_2$ and $V_1 > V_2$ in
LiZn$_2$Mo$_3$O$_8$, we think the relevant regime should be 
$ \mathbb{K}_1 \gg \mathbb{K}_2 $.
 
Because a strong PCO causes a strong modulation in the bond energy, 
the values of $\mathbb{K}_1$ and $\mathbb{K}_2$ for the resonating hexagon would be modified
from the effective Hamiltonian that is obtained from the perturbative analysis.
Likewise, the spin exchange in the resonating hexagon is enhanced from its perturbative value.
So we expect this treatment is a good approximation to understand the local spin physics. 

The Hilbert space of the Hamiltonian $H_{\text p}$ is spanned by the 
electron states that are labelled by the positions and the spins of the 
3 resonating electrons. Because the electrons are separated from each other
by one lattice site, the Hilbert space for the positions is quite limited. 
There are a total of 16 states labelled by
$\{ |\alpha\beta\gamma\rangle_{\text{A}} \equiv |2\alpha,4\beta,6\gamma \rangle,
|\alpha\beta\gamma \rangle_{\text{B}} \equiv | 1\alpha,3\beta,5\gamma \rangle  \}$
with $\alpha,\beta,\gamma = \uparrow, \downarrow$.  
Since the local Hamiltonian $H_{\text p}$ commutes with 
the total electron spin 
${\bf S}_{\text{tot}}$ and $S^z_{\text{tot}}$, 
we can use 
$\{ {\bf S}_{\text{tot}},  S^z_{\text{tot}}\}$ to label
 the states. 
For both A and B electron configurations, from
the spin composition rule for 3 spins ($\frac{1}{2}\otimes \frac{1}{2} 
\otimes \frac{1}{2} = \frac{1}{2} \oplus \frac{1}{2} \oplus \frac{3}{2} $), 
we have 2 pairs of ${\bf S}_{\text{tot}} =1/2$ states.

\begin{figure}[t]
{\includegraphics[width=8cm]{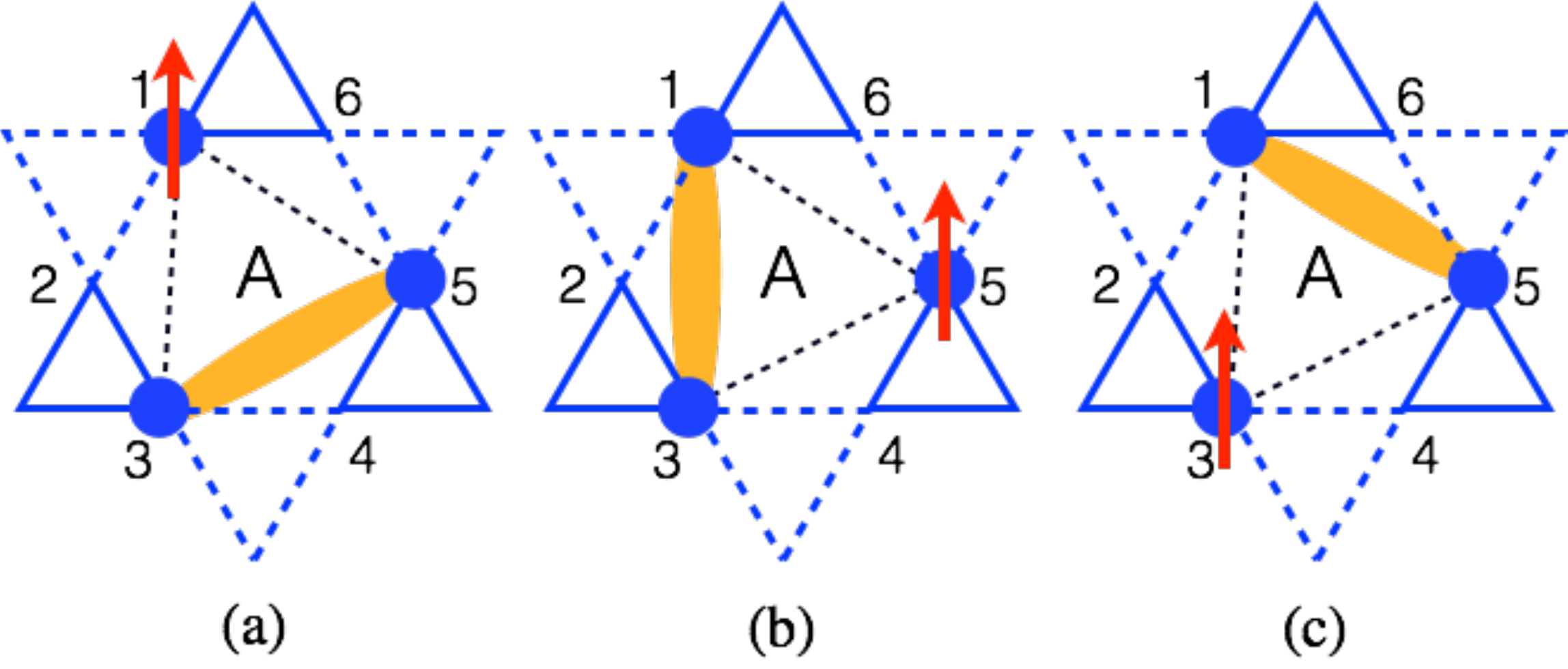}}
\caption{(Color online.) 
Three singlet positions that are related by the 3-fold rotation.  }
\label{fig13}
\end{figure}

The states with $S_{\text{tot}} = {3}/{2}$ are not favored by the AFM exchange. 
We find that when 
 $J > \frac{2}{3} (\mathbb K_1+ \mathbb K_2 -\sqrt{ \mathbb K_1^2 -\mathbb K_1
 \mathbb K_2 +\mathbb K_2^2  } ) 
=\mathbb K_2 - \frac{\mathbb K_2^2}{4 \mathbb K_1} + \mathcal{O} (\mathbb K_2^3)$,
the local ground states are 4 symmetric states with $S_{\text{tot}} =1/2$. 
Here, symmetric state refers to the symmetric linear superposition of 
spin states in A and B configurations. This is because the 
3-electron $\mathbb K_1$ and $\mathbb K_2$ hopping terms hybridize
the A and B configurations and favor symmetric superposition rather 
than antisymmetric superposition. 
The local 4-fold degeneracy can be effectively 
characterized by 2 quantum numbers ($s^z,\tau^z$) with  
$s^z = \pm \frac{1}{2}$ and $\tau^z = \pm \frac{1}{2}$.
 $s^z$ refers to the total spin 
$s^z \equiv S^z_{\text{tot}} = \pm \frac{1}{2}$. 
We also introduce a pseudospin-${1}/{2}$ 
operator $\boldsymbol{\tau}$ whose physical meaning is explained below.
The wavefunctions of the four $| {\tau^z s^z} \rangle$ states are 
given by (to the order of $\mathcal{O} ({\mathbb{K}_2}/{\mathbb{K}_1})$),
\begin{eqnarray}
|{\uparrow \uparrow }\rangle &=& \frac{1}{2} \big[
 |{\uparrow\uparrow\downarrow} \rangle_{\text A} 
 - |{\uparrow\downarrow\uparrow} \rangle_{\text A} 
+ | {\downarrow \uparrow\uparrow}\rangle_{\text B}
- |{\uparrow\uparrow\downarrow} \rangle_{\text B} 
  \big]
  \label{equu}
\\
 |{\downarrow\uparrow} \rangle &=& \frac{\sqrt{3}}{6} \big[2 |{ \downarrow \uparrow\uparrow}
 \rangle_{\text A}
 - | { \uparrow\downarrow \uparrow} \rangle_{\text A} - |  {\uparrow \uparrow\downarrow}
  \rangle_{\text A}
  \nonumber \\
  && \quad\,  + 2|  {\uparrow\downarrow\uparrow} \rangle_{\text B} - | { \uparrow \uparrow\downarrow} \rangle_{\text B}
   - | {\downarrow \uparrow\uparrow} \rangle_{\text B} \big]
     \label{eqdu}
\end{eqnarray}
and $|{\uparrow \downarrow} \rangle$, $|{\downarrow \downarrow} \rangle$ 
are simply obtained by a time-reversal operation. 

What is the physical origin of this local 4-fold degeneracy? 
Clearly, the 2-fold degeneracy of $s^z =\pm 1/2$ arises from 
the time-reversal symmetry and the Kramers' theorem. 
The remaining 2-fold degeneracy comes from the point group symmetry of the resonating hexagon. 
This is easy to see if we 
freeze the positions of the 3 electrons. To be concrete, let us fix the 
electrons to the sites 1,3,5 in Fig.~\ref{fig13}. To optimize the
 exchange interaction, 2 electrons must form a spin singlet, 
 which leaves the remaining electron
as a dangling spin-1/2 moment. As shown in Fig.~\ref{fig13}, 
this singlet can be formed 
between any pair of the electrons and the different locations of the spin singlet 
are related by the 3-fold rotation. Even though there seems
to be 3 possible singlet positions, only 2 of them are linearly independent, which
 gives to the 2-fold $\tau^z$ degeneracy which survives even 
when the ring electron hopping is turned on. 
As a result, the pseudospin $\boldsymbol{\tau}$ is even under time-reversal 
and acts on the space of the singlet position or equivalently the dangling spin position. 
In fact, the two states in Eqs.~\eqref{equu} and \eqref{eqdu} comprise the E irreducible 
representation of the C$_{3v}$ point group. 

Since the spin ${\bf s}$ (pseudospin $\boldsymbol{\tau}$) 
is odd (even) under time reversal symmetry, when the external magnetic field 
is applied to the system, only the spin ${\bf s}$ ($s=1/2$) couples to the 
magnetic field. Therefore, the three electrons in the resonating hexagon 
behave like one spin $s=1/2$ in the magnetic field. This is how the Curie
constant with the PCO becomes 1/3 of the Curie constant without the PCO.

\section{Levin-Wen's variational dimer wavefunction approach} 
\label{asec3} 

Here we explain the string mean-field theory in Sec.~\ref{sec3B} in details. 
To solve the combined 
Hamiltonian of $\bar{H}_{\text{s}}$ and $\bar{H}_{\text{ring}}$ 
self-consistently, we obtain the effective spinon hoppings by evaluating 
the boson or rotor hopping amplitudes with respect to the variational dimer
ground state $\Psi(\{z_i \})$,
\begin{eqnarray}
\langle L^+_{\mu} ( {\bf R}) L^-_{\nu} ({\bf R}) \rangle & = &
\langle L^+_{\mu} ( {\bf R}) \rangle \langle L^-_{\nu} ({\bf R}) \rangle
\end{eqnarray}
where $\mu,\nu$ label the sublattices.
We also evaluate the parameter $M_{ijklmn}$ against
the spinon hopping Hamiltonian. Using the Wick theorem, we have
\begin{eqnarray}
M_{ijklmn} &=& \quad
\sum_{\alpha\beta\gamma} 
\langle f^\dagger_{i\alpha} f^{\phantom\dagger}_{j\alpha} \rangle
\langle f^\dagger_{k\beta} f^{\phantom\dagger}_{l\beta} \rangle
\langle f^\dagger_{m\gamma} f^{\phantom\dagger}_{n\gamma} \rangle
\nonumber \\
&& +  \sum_{\alpha} \langle f^\dagger_{i\alpha} f^{\phantom\dagger}_{n\alpha} \rangle
\langle f^\dagger_{k\alpha} f^{\phantom\dagger}_{j\alpha} \rangle
\langle f^\dagger_{m\alpha} f^{\phantom\dagger}_{l\alpha} \rangle
\nonumber \\
&& +  \sum_{\alpha} \langle f^\dagger_{i\alpha} f^{\phantom\dagger}_{l\alpha} \rangle
\langle f^\dagger_{m\alpha} f^{\phantom\dagger}_{j\alpha} \rangle
\langle f^\dagger_{k\alpha} f^{\phantom\dagger}_{n\alpha} \rangle
\nonumber \\
&&- \sum_{\alpha\beta}  \langle f^\dagger_{i\alpha} f^{\phantom\dagger}_{j\alpha} \rangle
\langle f^\dagger_{k\beta} f^{\phantom\dagger}_{n\beta} \rangle
\langle f^\dagger_{l\beta} f^{\phantom\dagger}_{m\beta} \rangle
\nonumber \\
&&- \sum_{\alpha\beta}  \langle f^\dagger_{i\alpha} f^{\phantom\dagger}_{l\alpha} \rangle
\langle f^\dagger_{k\beta} f^{\phantom\dagger}_{j\beta} \rangle
\langle f^\dagger_{m\beta} f^{\phantom\dagger}_{n\beta} \rangle
\nonumber \\
&&- \sum_{\alpha\beta}  \langle f^\dagger_{i\alpha} f^{\phantom\dagger}_{n\alpha} \rangle
\langle f^\dagger_{m\beta} f^{\phantom\dagger}_{j\beta} \rangle
\langle f^\dagger_{k\beta} f^{\phantom\dagger}_{l\beta} \rangle
\label{eq73}
 \\
&=& \chi_{ij}^3 + \frac{\chi_{ik}^3}{4} + \frac{\chi_{il }^3}{4}
       -\frac{3}{2} \chi_{ij} \chi_{ik} \chi_{il}, 
\label{eq74a}
\end{eqnarray}
where we have defined the $\chi$ variable as 
\begin{equation}
\chi_{ij} = \sum_{\alpha} \langle f^{ \dagger}_{i \alpha} f^{\phantom\dagger}_{j \alpha}  \rangle,
\end{equation}
and we have also used the three-fold rotational symmetry as well as the reflection symmetry of the 
hexagon in Eq.\eqref{eq74a} so that
\begin{eqnarray}
&&\chi_{ij} = \chi_{kl} = \chi_{mn}
\\
&&\chi_{ik} = \chi_{km} = \chi_{mi} = \chi_{jl} = \chi_{ln} = \chi_{nj} 
\\
&& \chi_{ik} = \chi_{jm} = \chi_{kn} .
\end{eqnarray}

\bibliography{ref}

\end{document}